\documentclass[usenatbib]{mnras}
\usepackage[utf8]{inputenc}
\usepackage{aas_macros}
\usepackage{amsfonts}
\usepackage{amsmath}
\usepackage{bm}
\usepackage{amssymb}
\usepackage{graphicx}
\usepackage[dvipsnames]{xcolor}
\usepackage{hyperref}
\hypersetup{colorlinks=true,allcolors=teal}
\usepackage{microtype}

\newcommand{\Nsamp}{\ensuremath{N_{\rm samp}}}

\newcommand{\avg}[1]{\ensuremath{\left\langle \,#1\, \right\rangle}}
\newcommand{\e}[1]{\ensuremath{{\rm e}^{#1}}}

\newcommand{\der}{\ensuremath{{\rm d}}}

\newcommand{\eqn}[1]{equation~\eqref{#1}}

\newcommand{\be}{\begin{equation}}
\newcommand{\ee}{\end{equation}}
\newcommand{\Cal}[1]{\ensuremath{\mathcal{#1}}}

\defcitealias{rw05}{RW05}

\title[Anisotropic simulated annealing]
      {A simulated annealing approach to parameter inference with expensive likelihoods} 

\author[A. Paranjape]{
Aseem Paranjape$^{1}$\thanks{E-mail: aseem@iucaa.in}
\\  
 $^1$ Inter-University Centre for Astronomy \& Astrophysics,
      Ganeshkhind, Post Bag 4, Pune 411007, India}

\begin{document}
\label{firstpage}
\pagerange{\pageref{firstpage}--\pageref{lastpage}}
\maketitle

\begin{abstract}
We present a new approach to parameter inference targeted on generic situations where the evaluation of the likelihood $\mathcal{L}$ (i.e., the probability to observe the data given a fixed model configuration) is numerically expensive. % 34
Inspired by ideas underlying simulated annealing, the method first evaluates $\chi^2=-2\ln\mathcal{L}$ on a sparse sequence of Latin hypercubes of increasing density in parameter (eigen)space. % 24
The semi-stochastic choice of sampling points accounts for anisotropic gradients of $\chi^2$ and rapidly zooms in on the minimum of $\chi^2$. % 22
The sampled $\chi^2$ values are then used to train an interpolator which is further used in a standard Markov Chain Monte Carlo (MCMC) algorithm to inexpensively explore the parameter space with high density, similarly to emulator-based approaches now popular in cosmological studies. % 43
Comparisons with example linear and non-linear problems show gains in the number of likelihood evaluations of factors of 10 to 100 or more, as compared to standard MCMC algorithms. % 27
As a specific implementation, we publicly release the code \textsc{picasa}: Parameter Inference using Cobaya with Anisotropic Simulated Annealing, which combines the minimizer (of a user-defined $\chi^2$) with Gaussian Process Regression for training the interpolator and a subsequent MCMC implementation using the \textsc{cobaya} framework. % 44
Being agnostic to the nature of the observable data and the theoretical model, our implementation is potentially useful for a number of emerging problems in cosmology, astrophysics and beyond.  % 28
% total 222 words
\end{abstract}


\section{Introduction}
\label{sec:intro}
\noindent
Statistical inference in the context of parametrized models of observable quantities forms a cornerstone in addressing a variety of scientific questions. Focusing on astrophysics and cosmology, in particular, the advent of targeted experiments and observatories \citep{Komatsu2010,Planck18-VI-cosmoparam,borucki16,ricker+14,hobbs+10,abbott+19-GWTC-1} and large surveys \citep{eBOSS20,abbott+22-DES-Y3cosmo} has led to unprecedented statistical precision being achieved on a number of observables spanning a large range of physical conditions and length scales in, both, the local and distant Universe. Ongoing and upcoming surveys and experiments are set to further enhance the statistical precision of measurements \citep{merloni+12,abazajian+16-CMBS4,Laureijs11,LSST+09,DESI+16,baker+19-LISA}, as well as open up new avenues for exploring the Universe \citep[see, e.g.,][]{koopmans+15}, thus firmly placing the onus on model builders to produce ever-more-accurate models of the Universe.

In this context, the use of semi-analytical techniques and full-fledged numerical simulations as theoretical machines has now become common in data analysis and statistical inference \citep{kitaura+16,elucidV,alam+21a,yuan+22a}. These numerical tools are typically much more time-consuming and computationally expensive than (much simpler, but less accurate) analytical models, as is needed in order to correctly capture a large number of non-linear and uncertain model elements. For example, it is now common to have to describe complex physical phenomena such as galaxy formation and evolution using semi-analytical or heuristic models involving between $10$-$20$ free parameters \citep{guo+11,henriques+15,alam+21b}.
Within the cosmology community, such situations have led to several parallel lines of investigation, all aimed at robust statistical inference using high-precision data. The recognition that not all aspects of a full numerical simulation may be relevant for modelling specific observables has led to fast but approximate simulation techniques (typically with free parameters calibrated using full simulations) being developed to model cosmic structure formation \citep{ss02,pinocchio,mfc11,pinocchio-reloaded,kh12,tze13,hmp15,fcsm16,cp18}. The requirement of precision of a few percent in predicting observables such as the power spectrum of cosmic matter fluctuations has led to the development of `emulators', which smoothly interpolate the observable in multi-dimensional parameter space using full numerical simulations performed on a sparse set of carefully chosen parameter locations \citep{heitmann+14,heitmann+16,lawrence+17,knabenhans+19}. More recently, several groups have begun exploiting machine learning techniques \citep[e.g.,][]{adb18,lppl18,villaescusa-navarro+21-CAMELS,ds21,delgado+22,schaurecker+22} to train artificial neural networks to reproduce the results of a full simulation at a fraction of the cost.
Similar lines of approach can also be found in the gravitational waves \citep{schmidt+21}, strong gravitational lensing \citep{hpm17} and exoplanets communities \citep{cranmer+21}.

The techniques listed above are typically optimized in a highly model- and/or observable-specific manner. It is therefore also useful to explore more general-purpose techniques for parameter inference that combine accuracy and speed. For example, the Derivative Approximation for LIkelihoods \citep[DALI,][]{sqa14} uses analytical expansions around a multivariate Gaussian (which would be the exact answer for a linear model with Gaussian data errors), to model the non-Gaussian posterior of non-linear models. Schemes invoking DALI have been used in recent analyses of specific data sets \citep{rs22,wlzs22}. To account for arbitrary levels of non-Gaussianity in the parameter posterior, however, accurate likelihood calculations are essential. In situations where each likelihood calculation requires an expensive (semi-)numerical simulation, this can make efficient sampling of parameter space -- using, e.g., Markov Chain Monte Carlo (MCMC) techniques -- problematic.

There have already been several attempts to circumvent these problems. E.g., \citet{yuan+22b} have explored a hybrid MCMC, in which an emulator for the 2-point correlation function observables is fed into a standard MCMC implementation, for analysing data from the BOSS survey. \citet{bcd22} have used a Metropolis-within-Gibbs scheme to accelerate the sampling of nuisance parameters when searching for gravitational waves in pulsar timing array (PTA) data. The distinction between fast (typically nuisance) and slow (typically physically interesting) parameters has also been exploited in the construction of standard MCMC samplers based on the Metropolis-Hastings algorithm \citep{lewis13}. 

In this work, we present a new algorithm to accelerate the exploration of parameter space for generic problems with a calculable likelihood (or, equivalently, a cost function). Unlike DALI-based schemes such as the one proposed by \citet{rs22}, we do not assume any specific functional dependence of the cost function on the parameters. %\footnote{Moreover, \citet{rs22} demonstrated their technique on a sparse subset of a \emph{pre-existing} MCMC chain of accepted parameter samples, which does not address the key issue of how to choose the sparse sample in the first place.}
Our method instead is inspired by the ideas underlying simulated annealing -- an established technique for approximate optimization which we recapitulate below -- and essentially involves a simultaneous minimization and sparse multi-dimensional exploration of the cost function landscape. Building on this, our algorithm then further sparsely samples the parameter space in the vicinity of the minimum, followed by the construction of an interpolation scheme to enable inexpensive evaluations of the cost function at arbitrary parameter values. This interpolator is finally fed into a standard MCMC implementation to perform rapid exploration of the parameter space. We will demonstrate that our method uses fewer cost function evaluations than a standard MCMC solution by factors ranging from $\gtrsim10$ to $\gtrsim100$ depending on the problem; this is clearly particularly useful when the cost function (or likelihood) is expensive to compute. Our implementation is released publicly in the form of a code \textsc{picasa} whose main features we describe later.

The paper is organised as follows. We describe our new method in section~\ref{sec:method}, followed by a description of the \textsc{picasa} framework in section~\ref{sec:picasa}. We present applications to some example linear and non-linear problems in section~\ref{sec:examples} and conclude with a discussion in section~\ref{sec:conclude}.

\section{Method}
\label{sec:method}

\subsection{Recap: simulated annealing}
\label{subsec:simann}
Simulated annealing \citep{kgv83} is a technique for locating the global optimum of a cost function that can depend on several parameters, useful for cases when the function evaluation is difficult. The idea involves starting a search at some finite `temperature', leading to high probabilities of accepting non-optimal parameters, and slowly decreasing this temperature as better solutions are explored. The technique therefore allows for a stochastic exploration of parameter space that can effectively `jump out' of local optima, unlike standard optimization algorithms such as gradient descent.

The original implementations of simulated annealing were closely connected to the Metropolis-Hastings algorithm \citep{metropolis+53,hastings70}, and probabilistically accept or reject parameters sequentially
(\citealp{kgv83}, although see \citealp{ds90,mf90} for a deterministic updating approach). The gradual decrease of the `temperature' of the system ensures that the probability for accepting unlikely parameter values steadily decreases as the algorithm progresses. Adjusting the rate of this decrease (or the annealing schedule) leads to some level of control over the amount of resources that will eventually be spent in the cost function evaluations.

Our new method, described in the subsequent sections, is conceptually inspired by the idea of a steadily decreasing temperature. In particular, the algorithm we describe below progressively zooms in on smaller and smaller regions of parameter space enclosing the global minimum of a given cost function. However, we replace the probabilistic accept/reject scheme with a stochastic sampling strategy that leads to multiple parameter values being simultaneously sampled and stored. This is useful when we interface our algorithm with a standard MCMC implementation using interpolation schemes for evaluating the cost function. We also introduce several additional features that enable the algorithm to efficiently find a path towards the global optimum in multi-dimensional parameter space.

\subsection{Anisotropic Simulated Annealing}
\label{subsec:ASA}
In the following, we will use $\chi^2(\mathbf{a})$ to denote the cost function in the $D$-dimensional space of parameters $\mathbf{a}$. Throughout, we interpret $\chi^2$ as 
\be
\chi^2(\mathbf{a}) \equiv -2\,\ln\Cal{L} + {\rm constant}\,,
\label{eq:chi2}
\ee
where $\Cal{L}\equiv p(\Cal{O}|\mathbf{a})$ is the likelihood, or probability density (in observable space) of observing the data set $\Cal{O}$ given a model with fixed parameters $\mathbf{a}$. For many applications of interest, when data errors are assumed to be (approximately) normally distributed with a parameter-independent covariance matrix, the constant can be assumed to remove the effect of determinant factors, so that $\chi^2(\mathbf{a})$ corresponds to the quantity usually minimized in least squares applications. This allows us to cleanly connect to standard Bayesian MCMC applications later.\footnote{For cases where the likelihood is significantly non-Gaussian (e.g., count statistics in cluster cosmology), the constant can be adjusted so as to ensure that the minimum value of $\chi^2$ remains close to the number of degrees of freedom.} We emphasize again that we are not assuming any specific dependence of $\chi^2$ on the model parameters $\mathbf{a}$ (e.g., unlike DALI, we are not assuming that $\Cal{L}$ is close to a multi-variate Gaussian in the parameters).

\subsubsection{Iterative sampling}
The anisotropic simulated annealing (ASA) algorithm is built around the idea that an approximate minimum of $\chi^2(\mathbf{a})$ can be obtained using a sparse stochastic sampling in parameter space, and then iteratively refined. With some assumption on parameter priors, the samples obtained over multiple iterations can then also be used to map out the parameter posterior distribution around the minimum of $\chi^2$, using appropriate $D$-dimensional interpolation.

To this end, the ASA algorithm first evaluates $\chi^2(\mathbf{a})$ on Latin hypercube samples generated in progressively zoomed regions of the $D$-dimensional parameter space, with progressively higher density, for some number of iterations. The zoomed region at iteration $n$ is centered on the evaluated minimum of iteration $n-1$, with a volume decided by the estimated width of $\Cal{L}=\e{-\chi^2/2}$ in each parameter direction at iteration $n-1$, and having the same number of samples \Nsamp. The shape of the zoomed region at iteration $n$ is allowed to respond to anisotropic gradients of $\chi^2$ by setting the parameter range in the $p^{\rm th}$ direction to the dispersion $\sigma_p^{(n-1)}$ estimated by treating $\Cal{L}$ defined by the points in iteration $n-1$ as a probability density along each parameter direction and numerically evaluating the corresponding variance: $\sigma_p^{(n-1)2}  = \int \der a_p\,\Cal{L}_{(n-1)}(a_p - \avg{a_p})^2/\int \der a_p\,\Cal{L}_{(n-1)}$, where $\avg{a_p} = \int \der a_p\,\Cal{L}_{(n-1)}\,a_p/\int \der a_p\,\Cal{L}_{(n-1)}$.\footnote{If the initial range of parameters is very broad and the sampling is not dense enough, the first few iterations may be dominated by points having exceedingly large $\chi^2$ values, which would incorrectly lead to the dispersion estimate being numerically zero. To avoid such situations, we artificially replace all values of $\chi^2 > 20$ with zero. The starting iterations in such cases then essentially respond to the user-specified width, while later iterations become increasingly (and correctly) dominated by points having $\chi^2$ values closer to the number of degrees of freedom.} The use of points only in iteration $n-1$, rather than in all previous iterations, means that the algorithm explores regions of $\Cal{L}$ progressively closer to its maximum. The choice of setting the width of subsequent iterations exactly equal to the dispersion calculated above effectively fixes the rate of decrease of the `temperature' of the annealing process.

Since the \Nsamp\ evaluations in each iteration are completely independent, they can be trivially parallelized over $N_{\rm proc}\geq1$ processors. Maximum efficiency is then clearly achieved when $N_{\rm proc}$ is a multiple of \Nsamp.

\subsubsection{Detecting edges}
\label{subsubsec:edges}
If the estimated minimum is too close to the edge of the parameter range in the $p^{\rm th}$ direction at any iteration $n$, the estimate of the dispersion along that direction may be inaccurate, leading to errors that could propagate into subsequent iterations. To avoid this, in such a situation the algorithm forces the \emph{width} of the parameter range in the $p^{\rm th}$ direction for iteration $n+1$ to be the same as in iteration $n$ for that direction, but centered on the updated estimate of the minimum. Thus, the algorithm can automatically `claw' its way outside ranges that are too narrow, in its search for the true minimum. 

The condition of being `too close' is parametrized by a variable $\epsilon_{\rm close}$. If the absolute difference between the location of the minimum and either edge of the current parameter range along the $p^{\rm th}$ parameter direction is less than $\epsilon_{\rm close}$ times the width of the current parameter range along that direction, then the minimum is declared `too close' to the edge. Unless specified, we set $\epsilon_{\rm close}=0.1$ in what follows.

\subsubsection{Convergence criterion}
The iterations are terminated when a pre-decided convergence criterion is met. Below, we use two criteria to define convergence:
\begin{enumerate}
\item[(a)] The values of the minimum $\chi^2$ between iterations $n-1$ and $n$ change by a relative amount smaller than $\epsilon_{\rm conv}$, i.e.,
\be
|\chi^2_{{\rm min},(n-1)}/\chi^2_{{\rm min},(n)} - 1| \leq \epsilon_{\rm conv}\,, 
\label{eq:ASAconv}
\ee
where $0<\epsilon_{\rm conv}<1$. We demand that this criterion be met over two successive iterations.
\item[(b)] The flow towards the minimum of $\chi^2$ is converging, i.e., the estimated dispersion along each parameter direction must have decreased from iteration $n-1$ to $n$.
\end{enumerate}

Convergence is declared to be achieved when both these criteria are simultaneously met.

\subsubsection{Escaping local minima}
In practice, it is possible for the above algorithm to become stuck in a local minimum, with the convergence criterion being properly met but at relatively large $\chi^2$. This is because, as we mentioned earlier, the rate of decrease of the annealing `temperature' in our case is fixed by our choice of how the width of subsequent iterations is made. This can sometimes lead to the temperature decreasing `too quickly' for a given data and model setup. (This is not surprising, since the annealing schedule in the original simulated annealing algorithm is also highly model specific.)

To detect such situations, our implementation in section~\ref{sec:picasa} performs a simple goodness-of-fit test using the evaluated minimum $\chi^2_{\rm min}$ and the number of degrees of freedom $d \equiv N_{\Cal{O}} - D$, where $N_{\Cal{O}}$ is the length of the data (or observable) vector \Cal{O}. We demand that the $p$-value assuming that the converged value $\chi^2_{\rm min}$ is $\chi^2$-distributed with $d$ degrees of freedom is larger than a certain threshold. If not, then the algorithm attempts to break out of the local minimum by starting a new iteration centered on the current minimum but with a large parameter range. 

In practice, we have found that such situations can also occur if the starting range is very broad and the sampling density is too low, in which case the algorithm can potentially get stuck in an erroneous local minimum created by numerical errors. In such situations, it can become difficult for the algorithm to settle onto a track towards the true minimum and, instead, resources are wasted in very unlikely regions of parameter space. It then becomes advisable to quickly detect such pathological situations and restart the algorithm with a narrower starting range and/or higher sampling density. This is included in our implementation below.

\subsubsection{Quadratic form estimate}
\label{subsub:quadform}
Once an acceptable minimum $\mathbf{a}_\ast^{\rm (eval)}$ over the sampled parameter points is located, we fit a positive-definite quadratic form $Q(\mathbf{a})$ in its vicinity:
\be
Q(\mathbf{a}) = \chi^2_\ast + (\mathbf{a} - \mathbf{a}_\ast)^T\,C^{-1}\,(\mathbf{a} - \mathbf{a}_\ast)\,,
\label{eq:quadform}
\ee
where the scalar $\chi^2_\ast$, the $D$-vector $\mathbf{a}_\ast$ and the symmetric $D\times D$ matrix $C$ are obtained using a standard least squares criterion by minimizing $|Q(\mathbf{a}^{(i)}) - \chi^{2}_i|^2$ for a given sample of points $\{\chi^2_i,\mathbf{a}^{(i)}\}$.  In practice, we ensure that $C^{-1}$ (which would be the Fisher matrix for strictly linear models) is positive definite by starting with a small number of points to fit for $Q$ and gradually increase the sample size outwards from $\mathbf{a}_\ast^{\rm (eval)}$ until a positive definite $C^{-1}$ is obtained, inverting which gives us $C$. 

This fitting exercise has the dual advantage of not only refining the estimate for the true minimum (i.e., we use $\mathbf{a}_\ast$ rather than $\mathbf{a}_\ast^{\rm (eval)}$ as our best estimate), but also providing a zeroth order estimate for the errors on the best-fit (assuming broad priors) through the covariance matrix $C$. In section~\ref{subsec:picasa:ASA}, we also discuss how failures in estimating a positive definite $C$ can be handled. 
%In principle, this step could be refined using DALI-based fits, which we leave to future work.

\begin{figure*}
\centering
\includegraphics[width=0.45\textwidth]{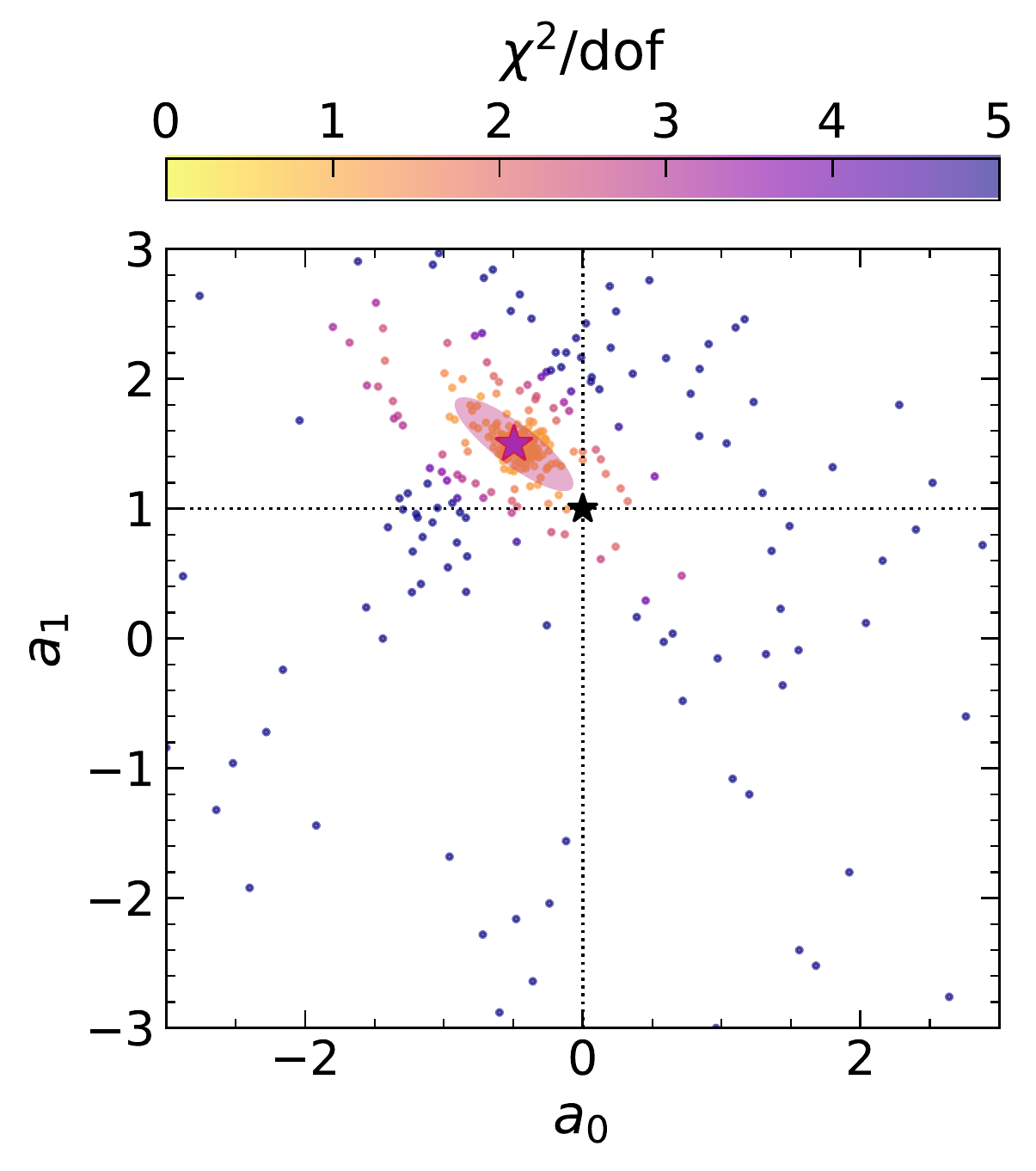}
\includegraphics[width=0.45\textwidth]{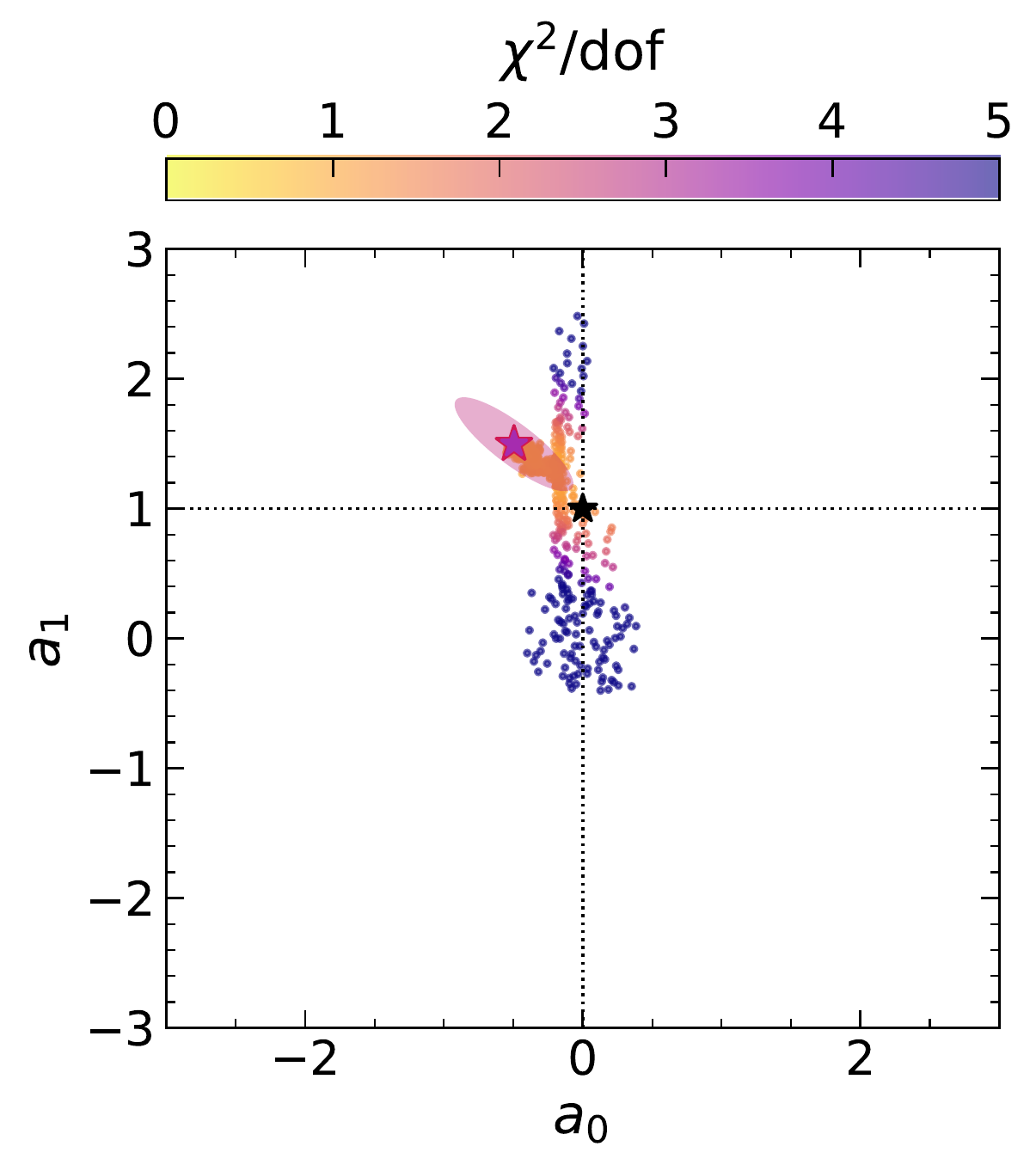}
\caption{{\bf ASA behaviour.} Distribution of sampled points for the 2-dimensional linear model discussed in section~\ref{subsec:visualdemo} assuming a wide (narrow) starting parameter range in the \emph{left (right) panel}, using $\Nsamp=50$. In each panel, the black star and dotted black lines indicate the true parameter values, the blue star and ellipse indicate the analytical best fit and $1\sigma$ confidence region (see equation~\ref{eq:visualdemo-analytical}), and the red star and ellipse (nearly indistinguishable from the analytical values) indicate the ASA best fit and $1\sigma$ confidence region. The sampled points are coloured by the corresponding value of reduced $\chi^2$, which shows how the ASA algorithm progresses towards regions of low $\chi^2$ in each case.}
\label{fig:visualdemo_scatter}
\end{figure*}

\subsection{Visual demonstration}
\label{subsec:visualdemo}

To demonstrate the basic behaviour of the ASA algorithm, consider some predicted scalar observable $y$ as a function of a $1$-dimensional predictor variable $x$, sampled at $N_{\Cal{O}}$ points $\{x_i\}$, $1\leq i\leq N_{\Cal{O}}$ (the algorithm is insensitive to this choice). The data $\{y_i\}$ are assumed to be normally distributed around the `truth' $y=f(x)$ with independent errors $\{\sigma_i\}$ (i.e., a diagonal covariance matrix with diagonal entries $\sigma_i^2$):
\be
y_i\sim\Cal{N}(f(x_i),\sigma_i^2)\,.
\notag
\ee
The cost function minimized by the ASA algorithm is then taken to be
\be
\chi^2(\mathbf{a}) = \sum_{i=1}^{N_{\Cal{O}}} \frac1{\sigma_i^2}\left(y_i - M(x_i;\mathbf{a})\right)^2\,,
\label{eq:example-chi2}
\ee
where $M(x;\mathbf{a})$ is the model function which depends on the $D$-dimensional parameter vector $\mathbf{a}$.

As a warm-up, and to visually display how the ASA algorithm locates the minimum of the cost function, consider the simple case $f(x)=x$ which we fit using the straight line $M(x;\mathbf{a}) = a_0 + a_1\,x$. For this exercise, we sampled the data at $11$ linearly spaced values between $0\leq x \leq 2$, assigning errors $\sigma_i = 0.5$ for each point. 
The minimization of $\chi^2$ (equation~\ref{eq:example-chi2}) can be performed analytically \citep[e.g.,][]{numerical-recipes} and gives the following best fit $\mathbf{a}$ and covariance matrix $C$ for the particular data realisation $\{y_i\}$ we used:
\begin{align}
a_0 &= -0.495253\,;\quad a_1 = 1.497192\,, \label{eq:visualdemo-analytical}\\
C_{00} &= 0.079545\,;\quad C_{01} = -0.056818\,;\quad C_{11} = 0.056818\,.
\notag
\end{align}
Notice that, although the data errors were assumed constant and independent (hence isotropic in data space), the resulting parameter covariance is not even diagonal, much less isotropic. The ASA algorithm must therefore explore a genuinely anisotropic parameter landscape. This is generically true for all the examples below.

To implement ASA, we must choose the starting range of parameters and the value of \Nsamp, which then also decides the starting sampling density. Fig.~\ref{fig:visualdemo_scatter} shows the sampled points when starting with a broad \emph{(left panel)} or narrow \emph{(right panel)} starting parameter range, using $\Nsamp=50$. In the first case, where the starting range comfortably enclosed the analytical values $(a_0,a_1)$ from \eqn{eq:visualdemo-analytical}, the algorithm quickly (within 2 zooms) centered itself close to the analytical values and converged to better than $0.1\%$ variation in  the minimum value of $\chi^2$ after 8 zoom iterations, i.e., $400$ $\chi^2$ evaluations in total. In the second case, where the analytical values were far outside the starting range, the algorithm took 8 zooms to discover the vicinity of the analytical values and subsequently converged after the 14th zoom level, i.e. $700$ $\chi^2$ evaluations in total. (The visualisation also demonstrates how the algorithm `clawed' its way to the minimum in this case; see section~\ref{subsubsec:edges}) Thus, a broader starting range (analogous to a higher starting `temperature') actually led to fewer overall evaluations for this simple case. The fitted quadratic form in the vicinity of the evaluated minimum $\chi^2$ for the first (second) case returns a best fit and covariance within one part in $10^9 (10^8)$ of the analytical values. This can also be seen from the excellent agreement between the red and blue stars and corresponding $1\sigma$ ellipses in Fig.~\ref{fig:visualdemo_scatter}. 

We also repeated this exercise with lower values of \Nsamp\ for the same starting parameter ranges for each case. For the first case (broad starting range), we find that \Nsamp\ as small as 10 still leads to excellent convergence centered on the analytical values with a correspondingly smaller number of function evaluations ($\simeq90$), and a similar recovery of the best fit and covariance matrix. The second case (narrow starting range) shows interesting behavior: while the algorithm itself converges increasingly farther from the analytical values as \Nsamp\ is decreased ($\gtrsim2\sigma$ away for $\Nsamp=10$), the resulting best fit and covariance matrix estimated from the fitted quadratic form remain within one part in $10^4$ of the analytical values. 

This demonstrates the interplay of \Nsamp\ and the starting parameter width for a simple 2-dimensional linear problem, and already suggests that, to ensure good sampling around the true minimum of $\chi^2$ in a generic situation, it is always useful to perform a few sparse iterations post-convergence, centered on the best-fit returned by the quadratic form fitting. Below we will see how this behaviour is affected by an increase in the dimensionality and/or non-linearity of the problem.

% \begin{figure}
% \centering
% \includegraphics[width=0.45\textwidth]{images/example_visualdemo_standard_data.pdf}
% \caption{{\bf ASA demonstration:} Green points with error bars show the data realisation drawn from the underlying true model $f(x)=x$ (dotted black line). Blue dashed line shows the analytical best fitting model. Solid red line with error band shows the ASA best fit along with the $68\%$ confidence region}
% \label{fig:visualdemo_data}
% \end{figure}

\subsection{Interpolating the cost function}
\label{subsec:interpolate}
Once the ASA algorithm converges, possibly along with additional iterations to improve the uniformity of sampling around the estimated minimum, we can use the resulting sample of points $\{\chi^2_i,\mathbf{a}^{(i)}\}$ to set up a $D$-dimensional interpolating function $\hat\chi^2(\mathbf{a})$. An accurate enough and computationally inexpensive enough $\hat\chi^2(\mathbf{a})$ can then be fed into a standard MCMC routine to explore parameter space.

Several choices of interpolating functions are in principle possible. We impose the requirements that the error in $\hat\chi^2(\mathbf{a})$ should be under control, and also that the choice of interpolation should be applicable to a wide variety of likelihood shapes (e.g., not necessarily close to Gaussian-shaped). We have found that a Gaussian Process Regression (GPR) defined with one of two kernels as discussed below works adequately, provided we employ an iterative training procedure. We give the details of our GPR usage below.

For an introduction to Gaussian Processes, we refer the reader to the excellent exposition in \citet[][hereafter, RW05]{rw05}. Any GPR is defined by the choice of its kernel, which defines the variance of the Gaussian probability density in function space which characterises the GPR. While the specific choice of kernel can be model-dependent, our focus on interpolating $\chi^2(\mathbf{a})$ rather than the prediction for the model itself suggests that a relatively small number of choices might suffice. In practice, we have found that either an anisotropic $D$-dimensional Gaussian kernel, or the so-called isotropic `rational quadratic' kernel, work adequately for a range of problems. Indeed, this is a distinct advantage of our method over the usual `emulator' approach which typically approximates the observable rather than the likelihood, and correspondingly requires highly specialised choices of kernel to accurately capture the full response to changes in parameter values.

Each GPR kernel is specified by a `hyper-parameter' vector $\mathbf{h}$: e.g., an anisotropic Gaussian kernel describing $\hat\chi^2(\mathbf{a})$ for a $D$-dimensional parameter space $\mathbf{a}$ requires a $D+1$-dimensional $\mathbf{h}$ to be specified (the dispersions in each parameter direction, along with an overall amplitude), while the isotropic rational quadratic kernel requires 3 parameters, namely, an amplitude, a dispersion scale and a shape parameter \citepalias[see chapter 4 of][]{rw05}. The values of the components of $\mathbf{h}$ are model-specific and need to be fixed by an appropriate comparison (or `training') with the provided sample $\{\chi^2_i,\mathbf{a}^{(i)}\}$. We use  Algorithm 2.1 of \citetalias{rw05} as implemented in Scikit-Learn, which maximises a (hyper-)likelihood function for a given training set, to set up an iterative training procedure as follows.
\begin{itemize}
\item We start with a small fraction of the available sample as the training set to estimate $\mathbf{h}$, using the ASA algorithm to explore hyper-parameter space.
\item We cross-validate this choice of $\mathbf{h}$ using the remaining samples and compute the $1$ and $99$ percentiles of the relative difference $\hat\chi^2(\mathbf{a})/\chi^2(\mathbf{a})-1$.
\item If the magnitudes of these percentiles are smaller than a pre-defined threshold $\epsilon_{\rm cv}$, then we declare the GPR as trained with the hyper-parameter $\mathbf{h}$. If not, we increase the size of the training set by a small amount ($20\%$, below) and repeat. 
\end{itemize}
If the GPR cannot be trained with more than, say, $80\%$ of the sample, 
%\AP{or when the cross-validation set size falls below 1000}, 
then the interpolation may not be accurate. This typically corresponds to situations where the $\chi^2$ sampling near sharp features is not fine enough, or when the overall sampling density is too low. In this case, one could either weaken the requirements for converged training (e.g., by increasing $\epsilon_{\rm cv}$ or/and thresholding on the 16, 84 percentiles rather than the extremes), or use a different kernel, or, failing all else, increase the sample size around the estimated minimum of $\chi^2$ and repeat. Our implementation in section~\ref{sec:picasa} checks for such failures and warns the user accordingly. In this regard, our technique is less robust than standard MCMC implementations which are mathematically guaranteed to sample the posterior distribution correctly. The gains we will see in section~\ref{sec:examples}, however, suggest that this decrease of robustness may be well worth the price for expensive likelihoods. We emphasize that the GPR training and its subsequent use in MCMC exploration is relatively inexpensive once a well-defined ASA sample is available (we provide some usage statistics later).

\section{\textsc{picasa} framework}
\label{sec:picasa}
We have implemented the ASA algorithm along with GPR training and subsequent sampling using standard MCMC, in the Python code \textsc{picasa} ({\bf P}arameter {\bf I}nference using {\bf C}obaya with {\bf A}nisotropic {\bf S}imulated {\bf A}nnealing). We use the publicly available \textsc{cobaya} framework \citep{tl19-cobaya,tl21-cobaya} to interface the ASA algorithm with MCMC sampling. We describe the main functionality of \textsc{picasa} in this section and discuss a few examples in section~\ref{sec:examples}.

\subsection{ASA implementation}
\label{subsec:picasa:ASA}
The ASA algorithm is implemented by a collection of methods in the Python class \texttt{ASAUtilities} which, by default, are called by the black-box methods defined in the class \texttt{PICASA}, but can also be directly accessed by the user. The user provides a dictionary (or `data pack') containing the function objects and data arrays required to calculate the cost function $\chi^2(\mathbf{a})$ for any specified parameter vector $\mathbf{a}$. In other words, both the model as well as shape of the likelihood are entirely user-defined, making our implementation generic. The data pack also requires the specification of \Nsamp\ (key \texttt{Nsamp}) and the starting range of parameter values (keys \texttt{param\_mins} and \texttt{param\_maxs}), along with a writable output file location (key \texttt{chi2\_file}).

Keeping in mind the discussion in section~\ref{sec:method}, the default implementation performs two basic passes over parameter space to locate the minimum $\chi^2$, followed by a nested sequence of independent `last iterations' that sample $\chi^2$ around its minimum. In detail, these steps proceed as follows:
\begin{enumerate}
\item In the first pass, the ASA algorithm starts with the user-specified parameter ranges and value of \Nsamp\ to crudely estimate the minimum of $\chi^2(\mathbf{a})$. This is done by setting $\epsilon_{\rm conv}=0.1$ (a relatively large value), allowing for quick convergence to a possibly inaccurate minimum.
\item Given this `coarse' sampling, the algorithm sets up a higher resolution second pass as follows. First, it attempts to perform the quadratic fit discussed in section~\ref{subsub:quadform}. If successful, the parameters are rotated into the eigenspace of the covariance matrix $C$ and the parameter ranges for the second pass are set to be $1\sigma$ wide in each eigen-direction around the estimated minimum. If a positive definite quadratic form cannot be obtained, the parameters are not rotated and the ranges for the second pass are set to be $10\%$ of the user-specified width, centered in this case on the currently evaluated minimum. We refer to this setup as the `minimizer mode' of the ASA algorithm.
\item The second pass of the ASA algorithm proceeds in this minimizer mode with the same \Nsamp, but with a more demanding convergence criterion $\epsilon_{\rm conv}$. By default this is set to be $\epsilon_{\rm conv}=0.001$ (i.e., only $<0.1\%$ variations in minimum $\chi^2$ are tolerated), but can also be specified by the user. In this mode, the algorithm uses high-density samples in small parameter volumes (`clawing' its way, if needed, towards the minimum; see section~\ref{subsubsec:edges}) to find a well-converged answer. 
\item If the minimizer mode is successful, a new estimate of the quadratic form is obtained using, by default, the full set of samples now available.\footnote{If a positive definite quadratic form cannot be obtained after the minimizer mode, the algorithm attempts to adjust by restarting the optimization procedure around the current minimum, using half the current starting range. The maximum number of such restarts is controlled by the data pack key \texttt{max\_opt\_iter} (default value $10$).} This is already expected to give a high-precision estimate of the true minimum. To facilitate further parameter exploration, the algorithm now proceeds to set up the `last iterations' sampling the vicinity of the true minimum.

This is done by performing a series of single ASA iterations, each using fixed hyper-rectangles in parameter eigenspace having widths ranging from $0.25\sigma$ to $\delta_\sigma\sigma$ in steps of $0.25\sigma$ along each eigen-direction. Here, the dispersion along each eigen-direction is determined by the eigenvalues of the covariance matrix $C$, and $\delta_\sigma$ (set by the data pack key \texttt{dsig}) is a user-defined constant. We have found that $\delta_\sigma\simeq5$-$6$ is sufficient for linear or nearly linear problems, while sufficiently non-linear problems may need $\delta_\sigma\gtrsim10$ (since the Gaussianised estimate of parameter error might not capture the true width of the non-Gaussian posterior). Working in eigenspace is very useful in ensuring efficient sampling when two or more parameters have strong degeneracies. 

The number of samples evaluated in each of these last iterations is controlled by a variable $f_{\rm last}$ (set by the data pack key \texttt{last\_frac}, with a default value $f_{\rm last}=0.05$), which specifies the fraction of the total sample size at the end of the minimizer mode that should be used to sample each iteration. Along with the value of $\delta_\sigma$, $f_{\rm last}$ can be adjusted by the user to ensure that the total number of samples is dominated by those in the last iterations, which helps the GPR training below to be performed efficiently. 
\item Steps (i) and (iii) above are further constrained by demanding that the total number of iterations in each not exceed 50 and 100, respectively. This ensures that, in cases where the algorithm gets pathologically stuck in local (numerically generated) minima, precious computational resources are not wasted in unlikely parameter regions. If either of these maximum iterations are exceeded, the code exits with a recommendation to the user to either decrease the width of the  initial sampling range, or increase \Nsamp, or both.
\item The user can optionally set the data pack key \texttt{parallel} to True, which invokes parallelization of the \Nsamp\ evaluations in any single iteration using the \texttt{multiprocessing} package. For optimal efficiency, the attribute \texttt{PICASA.NPROC} may be set to a multiple of $N_{\rm samp}$.
\item Finally, the code discards any samples where the value of $\chi^2$ is not finite. This can be utilised by the user to impose sharp priors, by simply requiring the model function to evaluate to infinity (\texttt{numpy.inf}) in undesirable or unphysical regions of parameter space.
\end{enumerate}
The end result of this exercise is a complete sample of finite $\chi^2$ values and parameter vectors, along with a final, more accurate estimate of the quadratic form close to the minimum, which can be fed into the GPR training procedure described in the next subsection. In cases where the problem is known to be nearly linear in the parameters, or where the location of the minimum is more important than its error, the ASA output at this stage may already provide all the required information. For this reason, the above steps are collected together in a single black-box method \texttt{PICASA.optimize}.

\subsection{GPR training}
\label{subsec:picasa:GPR}
We use the output of the ASA implementation above to train a GPR for interpolating $\chi^2(\mathbf{a})$ using the technique outlined in section~\ref{subsec:interpolate}. 
This is implemented by the method \texttt{PICASA.gpr\_train\_and\_run}. The choice of GPR kernel is specified by the data pack key \texttt{kernel} which can take values `rbf' (this is the default) for the anisotropic Gaussian kernel or `rq' for the isotropic rational quadratic kernel. 

The starting training set size is set to the minimum of 1000 and $5\%$ of the full sample, with a lower cap at 100. The training set is chosen by uniformly downsampling the full sample. The method invokes the ASA algorithm to optimize the GPR hyper-parameter vector $\mathbf{h}$, setting $\epsilon_{\rm conv}=0.005$ and $\epsilon_{\rm close}=0.2$. The starting training iteration employs ASA with $\Nsamp=50$ and a logarithmic hyper-parameter range centered on zero for each component of $\ln(\mathbf{h})$ with a width of $\pm\ln(2)$. 
While the cross-validation threshold $\epsilon_{\rm cv}$ (data pack key \texttt{cv\_thresh}) is exceeded, further training iterations are performed, using training sets successively larger by factors of $1.2$, $\Nsamp$ successively larger by factors of $1.15$ and ASA starting ranges successively centered on the current best estimate of $\ln(\mathbf{h})$ with widths successively larger by factors of $1.25$. Thus, the training iterations use gradually increasing training sets and explore gradually increasing regions of hyper-parameter space with gradually increasing \Nsamp. If the GPR training and cross-validation has not converged for a training set size larger than $80\%$ of the sample size, or if a maximum number of training iterations (by default 25) has been exceeded, the method exits with the last available hyper-parameter values and training size. In such cases, it is recommended to either alter the training thresholds and/or change the GPR kernel and/or increase the size of the original $\chi^2$ sample (say by performing additional last iterations, see also section~\ref{subsec:interpolate}). This introduces an element of problem-specific user intervention in the GPR training, which we comment on in section~\ref{sec:conclude}.

The user must supply a writable directory location (data pack key \texttt{chain\_dir}) which is used (among other things) to store the output of the training iterations, namely, the training size, values of hyper-parameter components and cross-validation percentiles. The final such outputs are used below for performing an MCMC run by feeding a trained GPR to the \textsc{cobaya} framework.

\subsection{MCMC exploration}
\label{subsec:picasa:mcmc}
We use the publicly available \textsc{cobaya} framework to perform MCMC sampling using a trained GPR to evaluate $\chi^2(\mathbf{a})$. We use the \texttt{Likelihood} class of \textsc{cobaya} to define the class \texttt{ASALike}, in which the method \texttt{ASALike.logp} calculates $-0.5\chi^2$ using the provided trained GPR. This setup is accessible to the \texttt{mcmc} sampler \citep{lewis13} available in \textsc{cobaya}, which we use for our MCMC exploration. The functionality for running \textsc{cobaya}'s \texttt{mcmc} sampler using the trained GPR is collected in the method \texttt{PICASA.run\_trained\_gpr}. This performs and stores the MCMC chains at the same writable location specified by the data pack key \texttt{chain\_dir} that was used by the GPR training above. We refer the reader to the \textsc{cobaya} documentation page for implementation details of the \texttt{mcmc} sampler and other usage.

We optimize the MCMC run by utilising the information collected during the ASA implementation, as follows.
\begin{itemize}
\item \emph{Priors} are chosen to be uniform in ranges $\pm\delta_\sigma\,\sigma$ along each parameter direction, centered on the best-fit parameter vector $\mathbf{a}_\ast$ obtained from quadratic form fitting on the full sample, where $\delta_\sigma$ is the same variable specified by the data pack key \texttt{dsig} (see section~\ref{subsec:picasa:ASA}). Note that the ASA optimization used $\delta_\sigma$ in eigenspace, in contrast to the parameter space priors here. 
\item The \emph{reference point} (starting location of each chain) is set equal to $\mathbf{a}_\ast$. 
\item The \emph{proposal covariance matrix} is initialised to be a small factor (default 0.1, set by the data pack key \texttt{prop\_fac}) times the best-fit covariance matrix $C$ from the quadratic form estimate. 
\end{itemize}
The \texttt{mcmc} sampler is run in its standard mode where it learns the proposal covariance matrix dynamically. The user can adjust the MCMC stopping criterion (data pack key \texttt{Rminus1\_stop}, default value 0.005). We emphasize that, since the MCMC sampling uses the trained GPR to estimate $\chi^2(\mathbf{a})$, rather than performing expensive actual calculations of $\chi^2(\mathbf{a})$, a demanding value of \texttt{Rminus1\_stop} such as 0.005 is not computationally challenging. Rather, it is the ASA parameter $\epsilon_{\rm conv}$ (accessible through the optional data pack key \texttt{eps\_conv} which specifies the value of $100\,\epsilon_{\rm conv}$) that behaves analogously in \textsc{picasa} to \texttt{Rminus1\_stop} in a standard MCMC run.

\begin{figure*}
\centering
\includegraphics[width=0.45\textwidth]{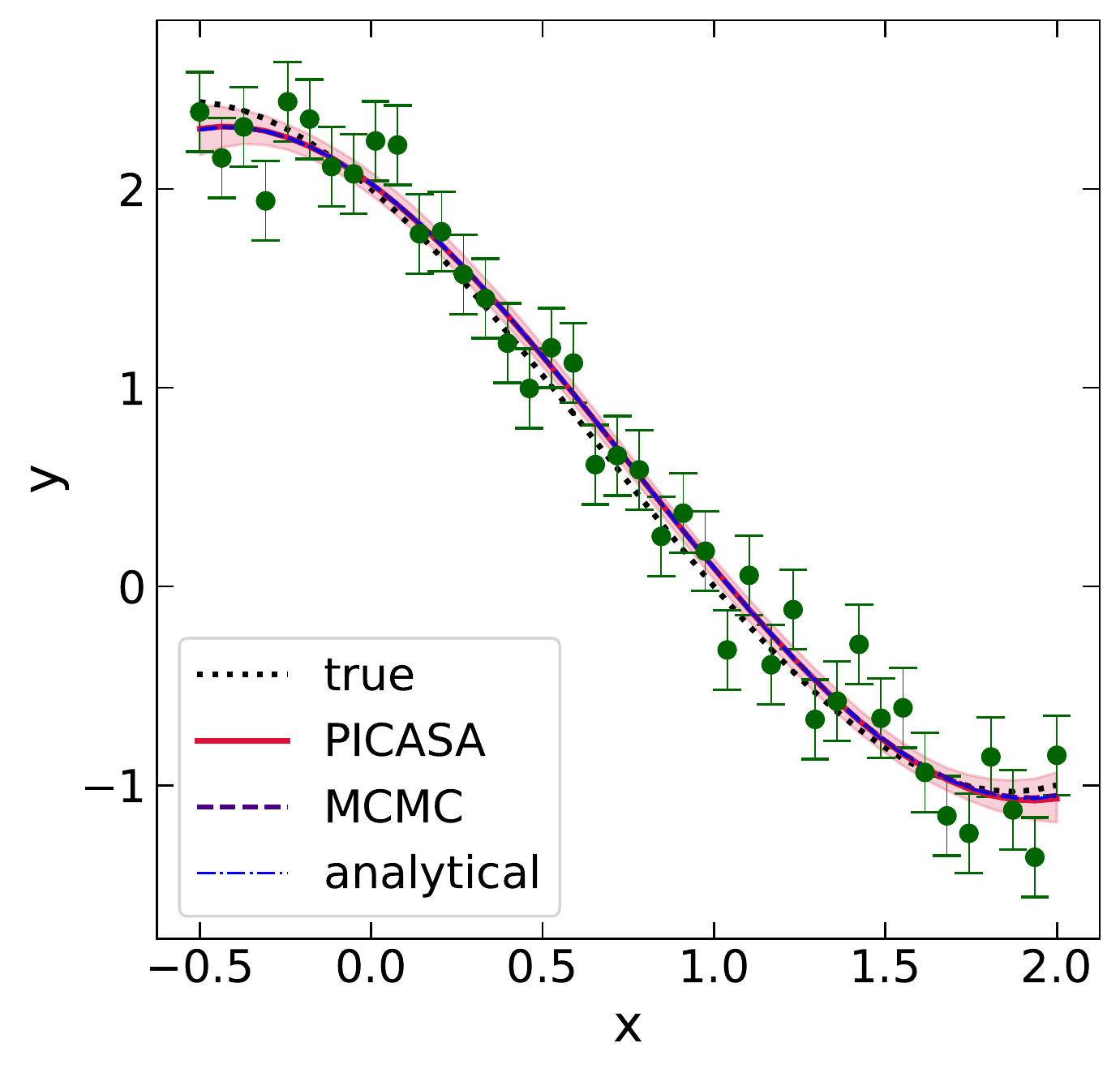}
\includegraphics[width=0.45\textwidth]{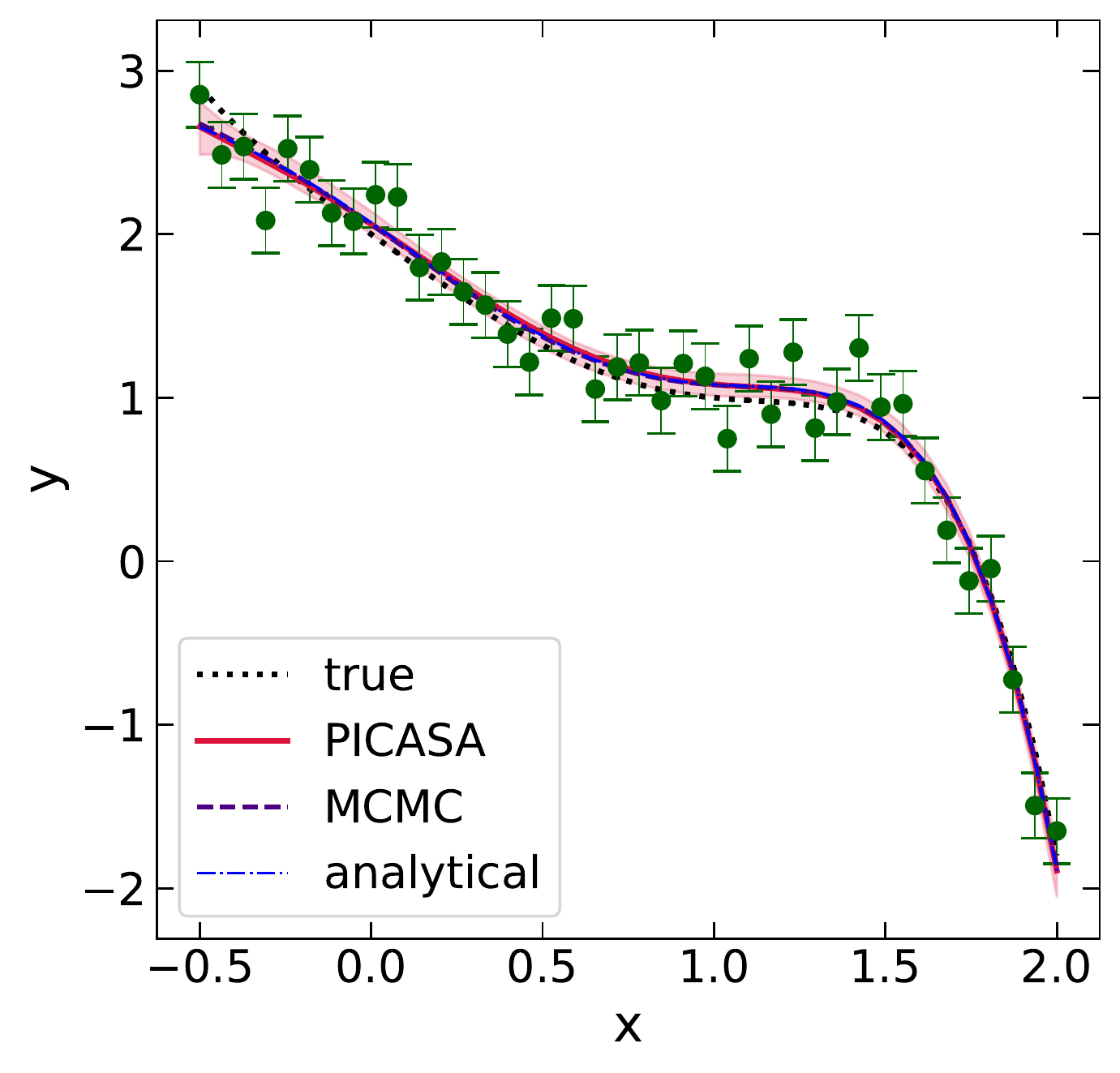}
\caption{{\bf PICASA versus MCMC: Linear models.} Recovery of a polynomial of degree $P=3$ ($P=5$) in the \emph{left (right) panel} using \textsc{picasa} (solid red curve with error band) and a standard MCMC implemented using \textsc{cobaya} (dashed purple curve). In each panel, green points with errors show the data realisation, dotted black curve shows the true model and dashed blue curve shows the analytical best fitting polynomial of degree $P$ (i.e., the equivalent of equation~\ref{eq:visualdemo-analytical} for the respective polynomial). See also Figs.~\ref{fig:poly_deg3_result} and~\ref{fig:poly_deg5_result}.}
\label{fig:poly_data}
\end{figure*}

For the user's convenience, we have collected together all aspects of this parameter inference framework in the single black box method \texttt{PICASA.mcmc}. This calls \texttt{PICASA.optimize} (if needed, repeatedly) to obtain accurate estimates of $\mathbf{a}_\ast$ and $C$, then calls the GPR training and MCMC running routines described above to produce and store a full MCMC sample. 
We have also provided some convenience flags using the data pack keys \texttt{no\_cost},  \texttt{read\_file\_mcmc} and \texttt{no\_train} that function as follows. 
\begin{itemize}
\item If \texttt{no\_cost} is True, the user must provide a single callable function that calculates $\chi^2$, rather than separate data and error arrays along with model and cost function callables. This allows for easier integration with pre-existing codes set up for MCMC sampling.
\item If \texttt{read\_file\_mcmc} is set to a valid file path containing $\{\chi^2_i,\mathbf{a}_i\}$ samples, then the ASA optimization step is skipped by \texttt{PICASA.mcmc} and the code directly proceeds to GPR training after reading the specified file. This is useful if the outputs of several, separate last iterations have been manually collected by the user.
\item Finally, if \texttt{no\_train} is True, then \texttt{PICASA.mcmc} assumes that a trained GPR has been specified at the default locations determined by the key \texttt{chain\_dir}, hence skips the GPR training and directly proceeds to the MCMC run. Combined with setting \texttt{read\_file\_mcmc}, this is particularly useful for restarting stuck chains, changing the stopping criterion, etc. 
\end{itemize}
We refer the reader to the \textsc{picasa} repository page listed under data availability below for further usage details.

\section{Examples}
\label{sec:examples}
We now turn to a more detailed comparison of the performance of the ASA algorithm and its integration with MCMC sampling using \textsc{picasa}, as compared to standard MCMC sampling. For simplicity, we use the same data structure discussed in section~\ref{subsec:visualdemo} (see the text around equation~\ref{eq:example-chi2}). Throughout, whenever discussing standard MCMC results, we discard only $1000$ accepted steps as burn-in. While this appears overly generous when comparing the total number of likelihood calls with \textsc{picasa}, it is offset by the fact that we will use a rather stringent convergence criterion for MCMC chains in all our comparisons. \textsc{picasa} does not suffer from burn-in, so that the total number of $\chi^2$ evaluations is relatively easy to track. Throughout, to define the standard MCMC, we will use the \texttt{mcmc} sampler \citep{lewis13} in the \textsc{cobaya} framework \citep{tl19-cobaya,tl21-cobaya}. We leave a comparison with other popular MCMC frameworks such as \textsc{emcee} \citep{f-mhlg13} or \textsc{zeus} \citep{kb20} to future work, although we expect similar results to hold.

\subsection{Linear models}
\label{subsec:linear}
We first consider polynomial forms for the `truth' $f(x)$ as prototypical of generic linear models: for a degree $P$ polynomial, we have
\be
f(x) = \sum_{p=0}^P\,\bar a_p\,x^p\,,
\label{eq:poly-def}
\ee
with $\bar{\mathbf{a}}=\{\bar a_p\}_{p=0}^P$ being the true values of the $P+1$ model parameters. We show results for the cases $P=3$ and $P=5$, leading to $4$- and $6$-dimensional parameter spaces, respectively. 
We use the following forms for the truth $f(x)$:
\begin{align}
P=3: &\quad f(x) = 2 - 3x/2 - x^2 + x^3/2 \label{eq:P3f(x)} \\
P=5: &\quad f(x) = 2 - 3x/2 + x^2/5 - x^3/5 + x^4 - x^5/2 \label{eq:P5f(x)}
\end{align}
We fit these using polynomial models of the respective degree $P$, i.e., we assume that the polynomial degree is known.\footnote{This is not a particularly restrictive assumption because, e.g., not knowing $P$ and fitting with a polynomial of degree $Q>P$ is equivalent to fitting a model with known degree $Q$ but having $\bar a_{P+1},\ldots, \bar a_{Q} = 0$.} In each case, we sample the data on 40 linearly spaced points in the range $-0.5 \leq x \leq 2$, with constant errors $\sigma_i=0.2$. The data realisation for $P=3$ ($P=5$) is shown in the \emph{left (right) panel} of Fig.~\ref{fig:poly_data}.

\begin{figure*}
\centering
\includegraphics[width=0.8\textwidth]{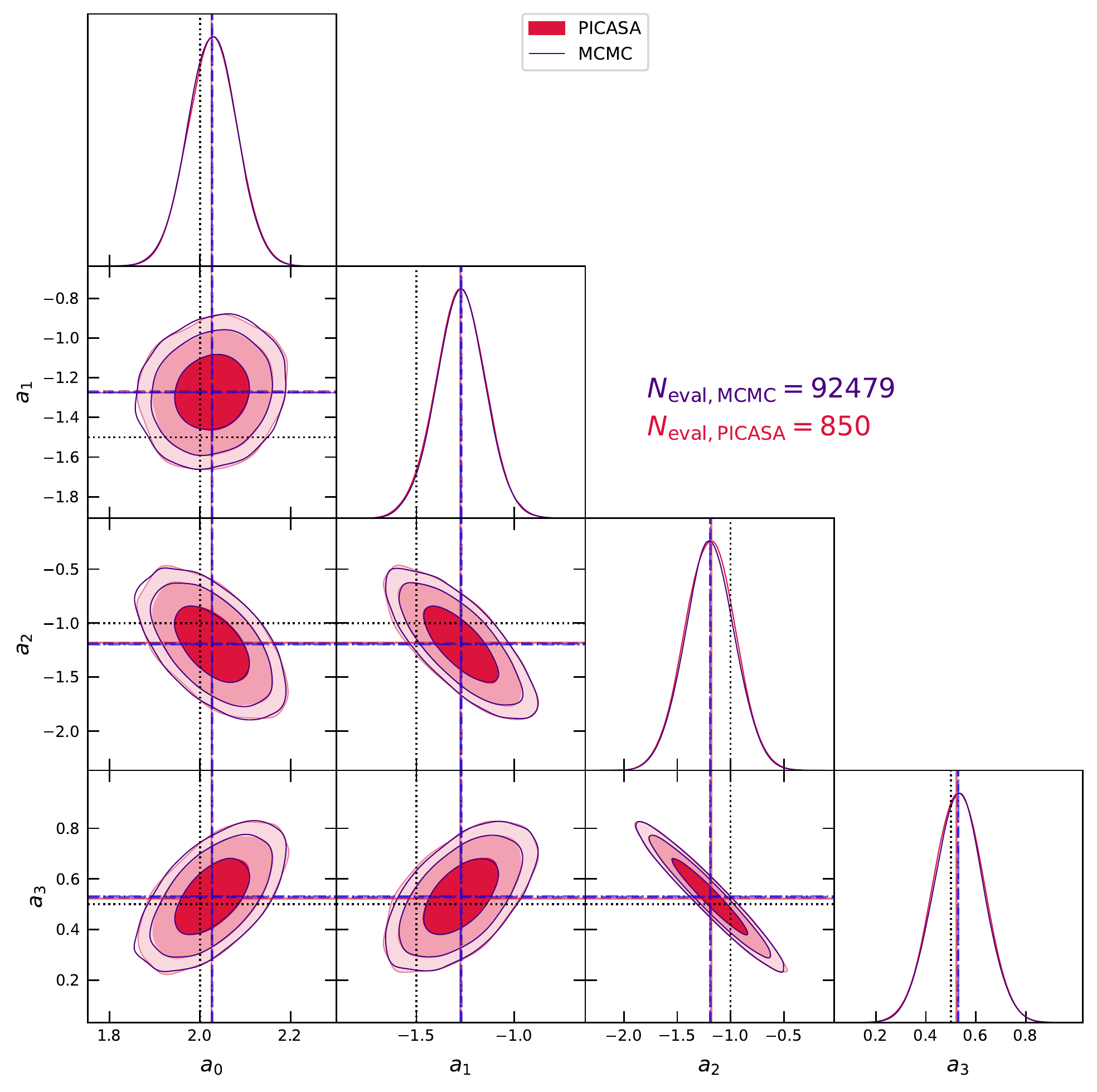}
\caption{{\bf Parameter constraints: cubic polynomial.} Corner plot showing pair-wise 1$\sigma$, 2$\sigma$ and 3$\sigma$ confidence contours for the parameters defining the cubic ($P=3$) polynomial in section~\ref{subsec:linear} (see equation~\ref{eq:P3f(x)}). Red filled (purple empty) contours show the results of \textsc{picasa} (standard MCMC). The correspondingly coloured vertical and horizontal lines indicate the respective best-fit values. The blue dashed vertical and horizontal lines indicate the analytical best-fit values. For comparison, the dotted black lines indicate the true parameter values from \eqn{eq:P3f(x)}. The labels indicate the total number of cost function evaluations in the respective runs. (For the MCMC, this is a slight underestimate, since we ignored the samples rejected during the burn-in phase, while the value for \textsc{picasa} is exact.) We see that the \textsc{picasa} result is essentially identical to that of the standard MCMC, despite using more than 100 times fewer evaluations, and also correctly accounts for the relatively strong degeneracy between $a_2$ and $a_3$.}
\label{fig:poly_deg3_result}
\end{figure*}

We use the full \textsc{picasa} framework with starting ranges as discussed below and the default set of last iterations out to $5\sigma$ in each parameter eigen-direction in each case (see section~\ref{sec:picasa} for details). The subsequent MCMC using a trained GPR (defined using the default anisotropic Gaussian kernel) is then compared with a standard MCMC implemented using \textsc{cobaya} with the \texttt{mcmc} sampler \citep{lewis13}. The latter is performed using uniform priors in the range $-5\leq a_p\leq 5$ and arbitrarily chosen reference points for each parameter, while the MCMC in \textsc{picasa} is initialised using the ASA output as described in section~\ref{subsec:picasa:mcmc}. Both sets of MCMC chains are run until the modified Gelman-Rubin index $R-1$ used by the \texttt{mcmc} sampler falls below $0.002$. %\AP{compare $R-1$ with $\epsilon_{\rm conv}$.}

\subsubsection{Cubic polynomial}
For the case $P=3$, we set $\Nsamp=30$ with starting ranges $-5\leq a_p\leq 5$ for each parameter for the ASA, and set a relatively low sampling for the last iterations using $f_{\rm last}=0.025$. 
The ASA optimization for the displayed example successfully completed using $850$ samples in total (i.e., including all last iterations). The iterative GPR training converged (with the 1,99 cross-validation percentiles being smaller than $0.01$) using $144$ of the $850$ samples. The resulting best fit and $68\%$ confidence interval in data space is shown by the solid red curve and error band in the \emph{left panel} of Fig.~\ref{fig:poly_data}. The corresponding $1\sigma$ and $2\sigma$ confidence regions in parameter space after MCMC sampling are displayed by the red contours in Fig.~\ref{fig:poly_deg3_result}.

\begin{figure*}
\centering
\includegraphics[width=0.9\textwidth]{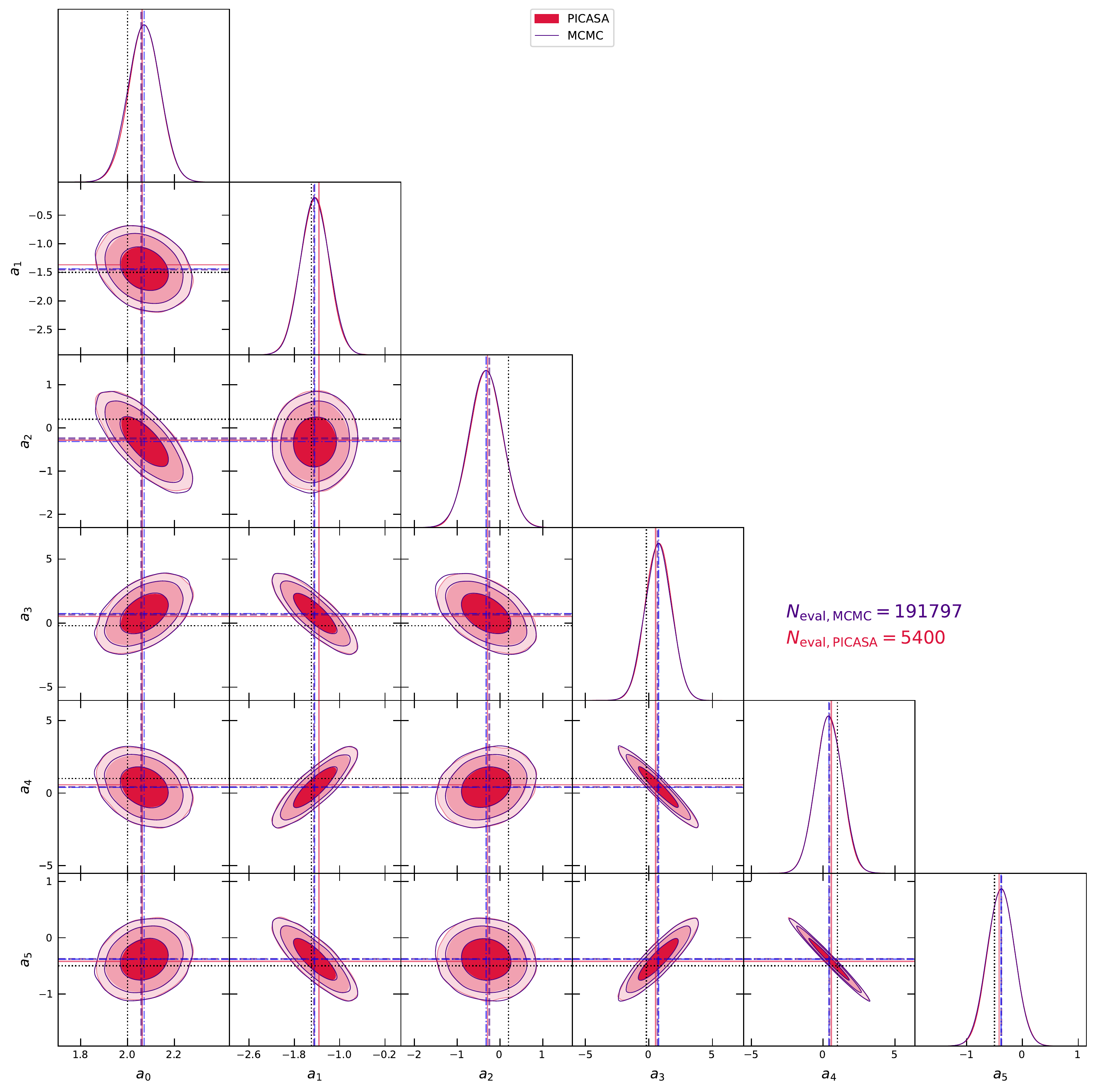}
\caption{{\bf Parameter constraints: quintic polynomial.} Same as Fig.~\ref{fig:poly_deg3_result}, showing results for the quintic polynomial ($P=5$) from \eqn{eq:P5f(x)}. We again see excellent agreement between \textsc{picasa} and the standard MCMC even at the 3$\sigma$ level despite strong parameter degeneracies, with \textsc{picasa} outperforming the standard MCMC by a factor $\sim35$. See text for a discussion of typical gains averaged over many runs.}
\label{fig:poly_deg5_result}
\end{figure*}

These results can be compared with the output of the standard MCMC, shown by the dashed purple curve in the \emph{left panel} of Fig.~\ref{fig:poly_data} and the purple contours in Fig.~\ref{fig:poly_deg3_result}. The analytical best fit is shown by the dot-dashed blue curve in the \emph{left panel} of Fig.~\ref{fig:poly_data}. We see excellent agreement between the \textsc{picasa}, standard MCMC and analytical results. The agreement between \textsc{picasa} and standard MCMC extends to the pairwise 3$\sigma$ contours, which is important for applications requiring the calculation of the Bayesian evidence. The key point to emphasize, though, is that the total number of likelihood evaluations $N_{\rm eval}$ in the standard MCMC exceeded $92000$, while \textsc{picasa} achieved its results with $850$ evaluations, a gain of a factor $\simeq108$. 
For a fair comparison, it is also important to consider that the total number of likelihood evaluations in both \textsc{picasa} and the standard MCMC can fluctuate from run to run. By repeating the standard MCMC sampling multiple times, we determined that the median and central $68\%$ region of the distribution of $N_{\rm eval}$ with the standard MCMC is $(58.9^{+20.0}_{-19.7})\times10^3$ for this data realisation of the $P=3$ model. A similar exercise for \textsc{picasa} with exactly the same setup as above but changing the random number seed for ASA sampling and subsequent steps gave $N_{\rm eval}=1365^{+796}_{-515}$, with $\sim 18\%$ of runs having optimization convergence failures leading to discarding $\Nsamp\times50=1500$ samples each (see section~\ref{subsec:picasa:ASA}). Thus, we see that gains of factors exceeding $50$ are typical when comparing \textsc{picasa} with standard MCMC, although \textsc{picasa} is less robust to the choice of starting parameter range. Indeed, simply decreasing the initial parameter range to $-2.5\leq a_p\leq 2.5$ for \textsc{picasa} led to $N_{\rm eval}=1080^{+721}_{-360}$ with a failure rate of only $3\%$.

\begin{figure*}
\centering
\includegraphics[width=0.45\textwidth]{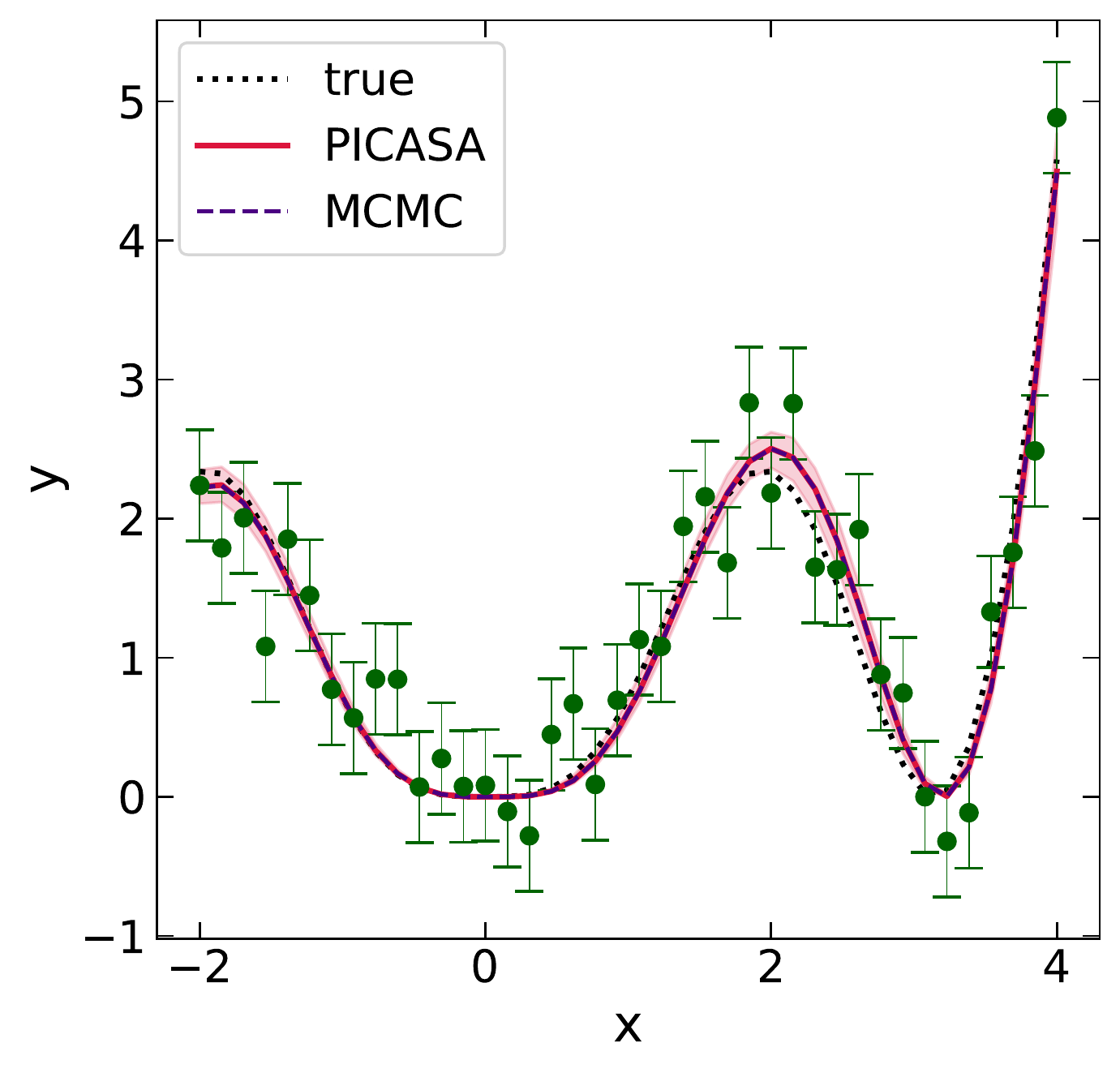}
\includegraphics[width=0.45\textwidth]{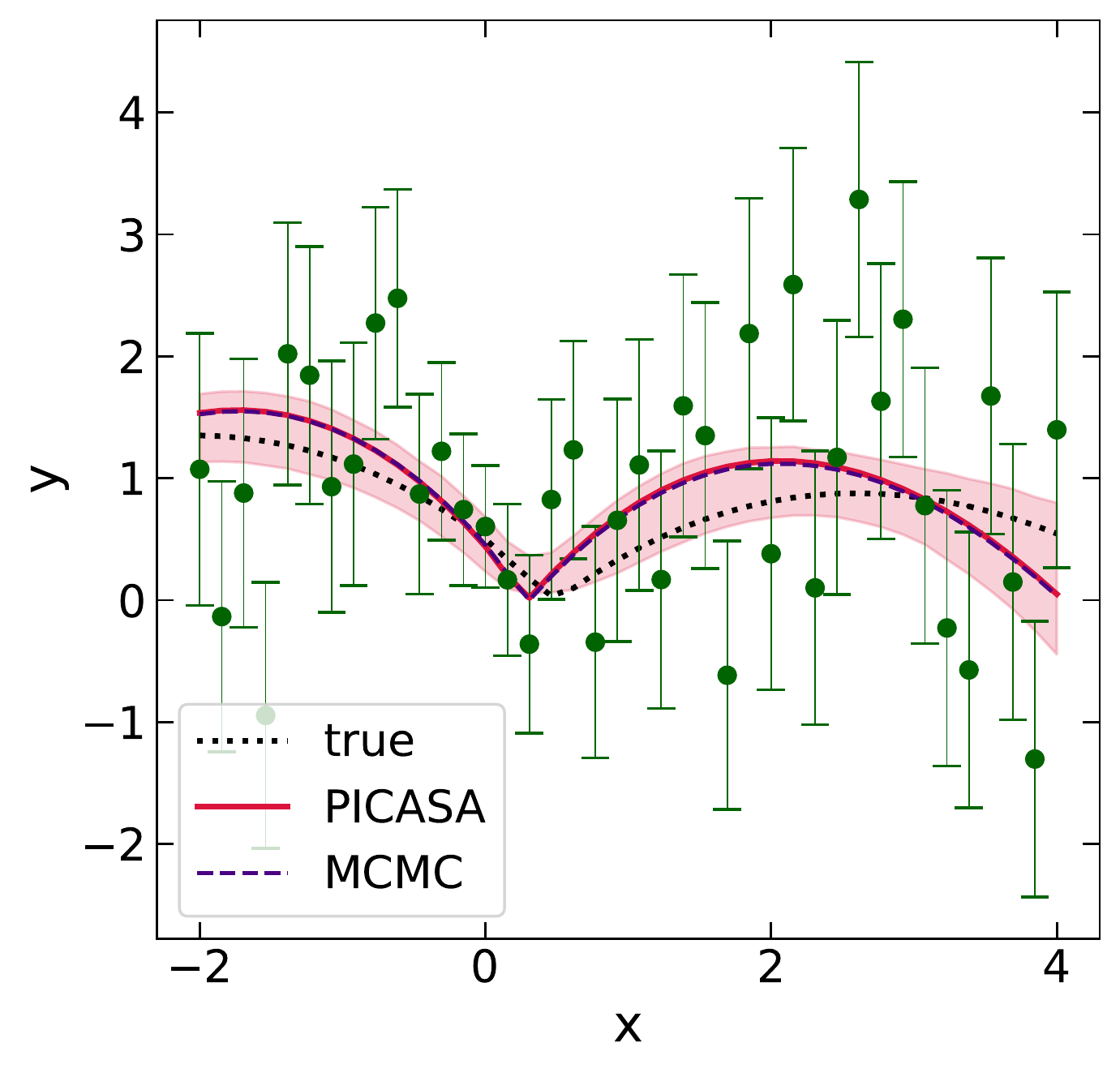}
\caption{{\bf PICASA versus MCMC: Non-linear models.} 
Same as Fig.~\ref{fig:poly_data}, showing the recovery of the non-linear models given by \eqn{eq:nonlin-truth-1} \emph{(left panel)} and~\eqn{eq:nonlin-truth-2} \emph{(right panel)}.  See also Fig.~\ref{fig:nonlin_result}.}
\label{fig:nonlin_data}
\end{figure*}

\subsubsection{Quintic polynomial}
We repeated a similar exercise for the $P=5$ case from \eqn{eq:P5f(x)}. For the example run displayed in the \emph{right panel} of Fig.~\ref{fig:poly_data}, we used $\Nsamp=100$ with a starting range $-3\leq a_p\leq 3$ and a relatively high sampling for the last iterations using $f_{\rm last}=0.1$, leading to a total sample size of $N_{\rm eval}=5400$, with GPR training convergence achieved using $388$ samples. Fig.~\ref{fig:poly_deg5_result} is formatted identically to Fig.~\ref{fig:poly_deg3_result} and shows the resulting comparison with the standard MCMC with uniform priors in the range $-5\leq a_p\leq 5$. Again, we see a comparable level of recovery of the analytical best fit using both \textsc{picasa} and standard MCMC, as well as very good agreement between \textsc{picasa} and MCMC even at 3$\sigma$, with the MCMC requiring $N_{\rm eval}\sim191000$, a gain by a factor of $\sim35$ when using \textsc{picasa}.

In statistical terms, in this case, a standard MCMC led to $N_{\rm eval} \simeq (129.1^{+43.1}_{-38.4})\times10^3$. For \textsc{picasa}, using a starting range of $-5\leq a_p\leq 5$ with \Nsamp\ as high as $300$ led to ASA optimization failure rates exceeding $50\%$, with $\Nsamp=500$ giving a lower failure rate of $27\%$ but a median $N_{\rm eval}\simeq41000$, suggesting that it is imperative to use a smaller starting range in this case. Using $-3\leq a_p\leq 3$ with $\Nsamp=100$ (the same as displayed in Fig.~\ref{fig:poly_deg5_result}) gave a failure rate of $40\%$ with $N_{\rm eval}\simeq(9.2^{+5.3}_{-2.3})\times10^3$, while $-2\leq a_p\leq 2$ with $\Nsamp=100$ gave a failure rate of $24\%$ with $N_{\rm eval}\simeq(9.5^{+4.1}_{-3.2})\times10^3$. We see that, despite high rates of optimization failure in this case, $\Nsamp\lesssim100$ gives a median $N_{\rm eval}$ of $\lesssim10^4$ when the run succeeds, compared to $\gtrsim10^5$ for the standard MCMC. We also emphasize that all successful optimization runs invariably led to quadratic forms that recovered the analytical best fit with very high precision (less than $0.01\%$ deviation in all cases).

Thus, we see that with increasing dimensionality, broad starting ranges lead to high rates of optimization failure in \textsc{picasa}, which fall as the starting range decreases in width, and \textsc{picasa} very effectively recovers the minimum of $\chi^2$ whenever the run succeeds. In practice, for example, starting with $-5\leq a_p\leq 5$ and $\Nsamp=30$ in our $P=5$ example above would almost certainly lead to a failed optimization, with $\simeq1500$ samples wasted. However, if the user then restarts \textsc{picasa} with a smaller starting range and a larger \Nsamp, convergence will typically be achieved within a few thousand evaluations, so that the overall gain compared to standard MCMC (even accounting for the wasted samples), would still be factors of $\sim10$-$50$.

\subsection{Non-linear models}
\label{subsec:nonlinear}
We next move on to models that are non-linear in the parameters $\mathbf{a}$, considering the following two examples for the `truth':
\begin{align}
f_1(x) &= |x|^{3/2}\,\sin^2(x)\,, \label{eq:nonlin-truth-1}\\ 
f_2(x) &= |x - 1/2|\,\cos(\sqrt{|x|/2})\,,
\label{eq:nonlin-truth-2}
\end{align}
which we sample in the range $-2\leq x\leq 4$. For the errors, we assume a constant value $\sigma(x)=0.4$ for $f_1(x)$ and $\sigma(x) = (1/2)\left[\tanh(|x|+1)/\tanh(1)\right]^3$ for $f_2(x)$. The resulting data realisations for $f_1$ and $f_2$ are respectively shown in the \emph{left} and \emph{right panel} of Fig.~\ref{fig:nonlin_data}.

We respectively fit these two data sets using the 3-dimensional models
\begin{align}
M_1(x;\mathbf{a}) &= a_0\,|x|^{a_1}\,\sin^2(x+a_2)\,, \label{eq:nonlin-model-1}\\ 
M_2(x;\mathbf{a}) &= a_0\,|x - a_1|\,\cos(\sqrt{a_2 |x|})\,.
\label{eq:nonlin-model-2}
\end{align}
In \eqn{eq:nonlin-model-2}, we have chosen to place the parameter $a_2$ under the square-root, along with a prior $a_2\geq0$, to avoid a perfectly degenerate maximum of the likelihood. Similarly, in \eqn{eq:nonlin-model-1}, we impose a prior $-\pi\leq a_2\leq \pi$ by ensuring that the model returns infinity outside this range (such values are discarded by the ASA sampling, see section~\ref{subsec:picasa:ASA}). The current implementation of ASA would otherwise always sample only one of the multiple, perfectly degenerate minima of $\chi^2$. We leave a more careful treatment of (nearly) degenerate minima of the cost function to future work.

\begin{figure*}
\centering
\includegraphics[width=0.48\textwidth]{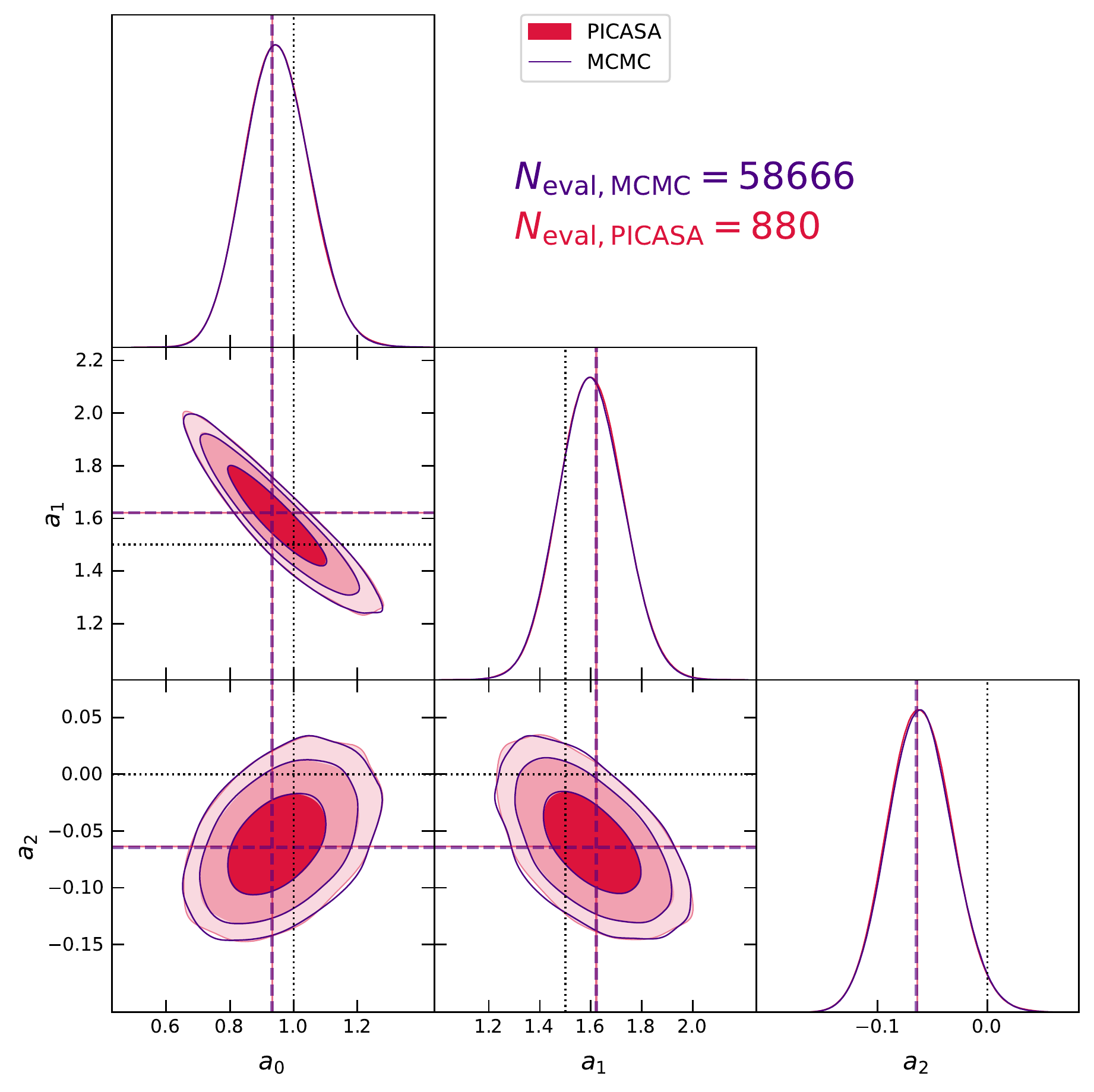}
\includegraphics[width=0.48\textwidth]{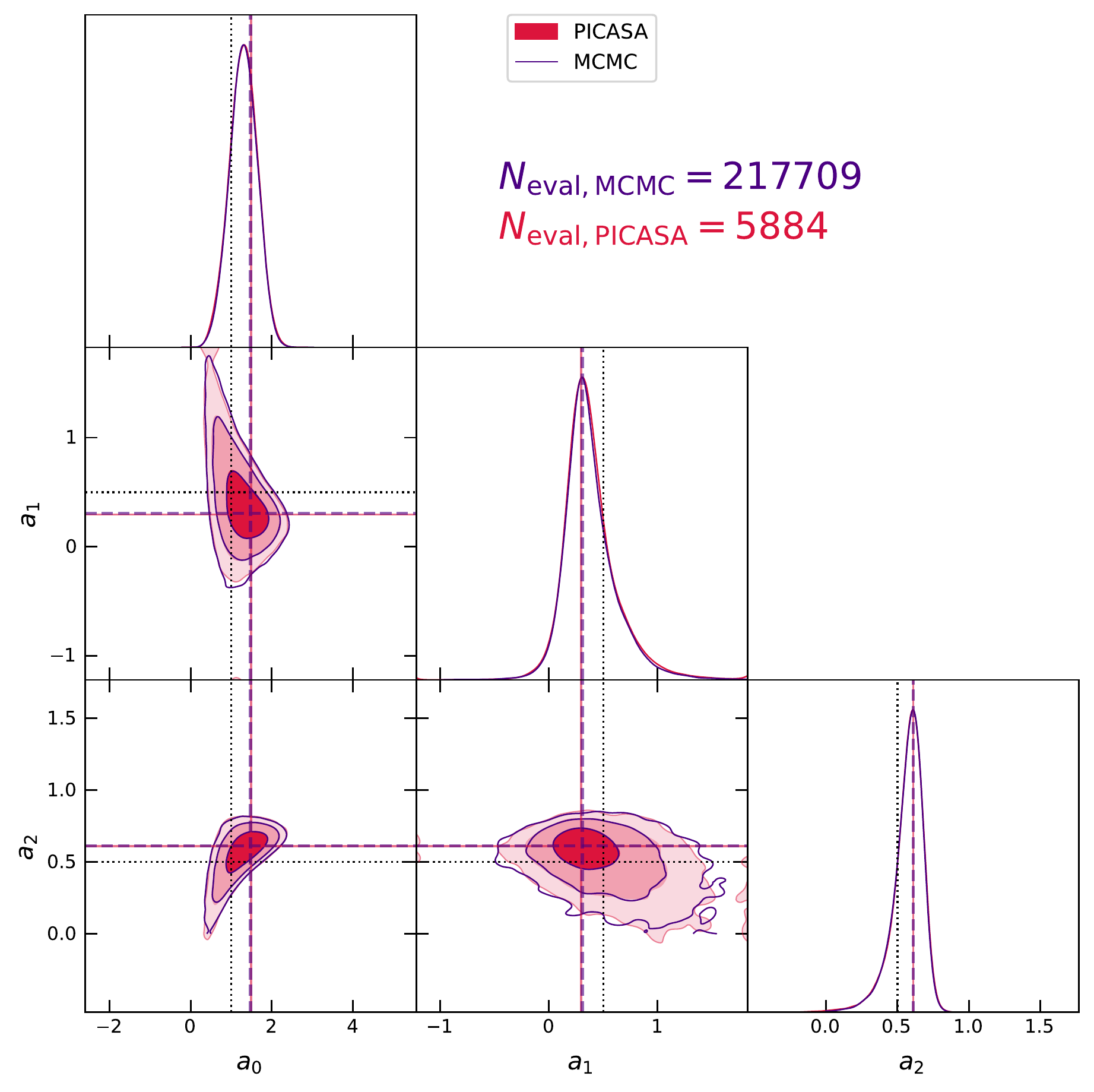}
\caption{Same as Figs.~\ref{fig:poly_deg3_result} and~\ref{fig:poly_deg5_result}, showing results for the non-linear models given by \eqn{eq:nonlin-truth-1} \emph{(left panel)} and~\eqn{eq:nonlin-truth-2} \emph{(right panel)}. We see excellent agreement between \textsc{picasa} and the standard MCMC, despite the high level of non-Gaussianity and parameter degeneracy, with \textsc{picasa} requiring $\sim35$-$65$ times fewer cost function evaluations. See text for a detailed discussion.}
\label{fig:nonlin_result}
\end{figure*}

The \emph{left} and \emph{right panels} of Fig.~\ref{fig:nonlin_result} are formatted identically to Fig.~\ref{fig:poly_deg3_result} and respectively show the MCMC results for recovering $f_1$ and $f_2$. In both cases, the \textsc{picasa} runs used starting ranges of $-1\leq a_p \leq 3$ for all parameters, with the same ranges also being used to set uniform priors for the standard MCMC runs. The MCMC convergence criterion was set to $R-1<0.002$ ($0.001$) for $f_1$ ($f_2$). While the \textsc{picasa} run for recovering $f_1$  succeeded with $\delta_\sigma=5$ and $f_{\rm last}=0.05$ (i.e., the default values), the recovery of $f_2$ used $\delta_\sigma=18$ and $f_{\rm last}=0.1$, and we found that $\delta_\sigma\lesssim10$ for this case led to highly truncated results and unstable GPR training. This is easy to understand in retrospect from studying the shapes of the 2$\sigma$ and 3$\sigma$ contours in the \emph{right panel} of Fig.~\ref{fig:nonlin_result}, which shows a highly non-Gaussian posterior. We emphasize that this choice of last iterations can be made \emph{after} the basic ASA optimization is complete, so it does not waste any evaluations. We also see that the 3$\sigma$ contours (particularly in the $a_1$-$a_2$ plane) are now noisier than in the other examples, which explains our choice of a more stringent MCMC convergence criterion. For the specific runs shown, \textsc{picasa} with $\Nsamp=30$ ($100$) led to recovery using $N_{\rm eval}= 880$ ($5884$) evaluations, a gain of a factor of $\sim65$ ($\sim35$) over the standard MCMC in recovering $f_1$ ($f_2$). For the GPR training for $f_1$ ($f_2$), we set the \texttt{robust\_convergence} key to False (True), the \texttt{cv\_thresh} key to $0.005\, (0.05)$ and the \texttt{kernel} key to `rbf' (`rq') (see section~\ref{subsec:picasa:GPR} for details). The GPR training then succeeded with $296$ ($2602$) samples. 

We also performed a statistical analysis of $N_{\rm eval}$ for each of the models, comparing the standard MCMC with \textsc{picasa}. 
For $f_1$, the median and central $68\%$ range of the distribution of $N_{\rm eval}$ for the standard MCMC, demanding $R-1<0.002$, was $N_{\rm eval}\simeq(32.2^{+11.5}_{-8.5})\times10^3$, while for \textsc{picasa} with the default $\delta_\sigma$ and $f_{\rm last}$ and $\Nsamp=30$, it was $N_{\rm eval}\simeq770^{+206}_{-120}$, with an optimization failure rate of $1\%$. Thus, \textsc{picasa} leads to typical gains of factors of $\sim40$ as compared to the standard MCMC in this case.
For $f_2$, the standard MCMC demanding $R-1<0.001$ used $N_{\rm eval}\simeq(104.4^{+66.6}_{-27.2})\times10^3$. With the $\delta_\sigma$ and $f_{\rm last}$  choices mentioned earlier, \textsc{picasa} with $\Nsamp=30$ led to $N_{\rm eval}\simeq 2460^{+984}_{-236}$ with an optimization failure rate of $5\%$, while increasing \Nsamp\ to $100$ (the displayed example) gave 
$N_{\rm eval}\simeq 6272^{+850}_{-645}$ and decreased the failure rate to $1\%$. 
% $N_{\rm eval}\simeq3500^{+462}_{-350}$ and decreased the failure rate to $1\%$. 
Thus, in this case, gains of factors of $15$-$40$ are typical. 

\subsection{Relative performance of ASA and GPR training}
Although \textsc{picasa} clearly outperforms a standard MCMC in terms of total number of cost function evaluations, in order for it to be a practical alternative to the latter, the computational expense of the GPR training step must also be understood. The main bottleneck in GPR training is the inversion of a $N_{\rm train}\times N_{\rm train}$ matrix, where $N_{\rm train}$ is size of the training set. Our iterative training procedure slowly increases $N_{\rm train}$ until convergence is achieved, making the total computational cost of GPR training highly problem-specific. In general, the GPR for likelihoods with features will of course be harder to train than for smooth likelihoods. 

The examples discussed above give us some idea of possible variations. We saw that the GPR for all but the non-linear $f_2$ model was successfully trained with a few hundred training samples, while $f_2$ used $\sim2600$ samples. In terms of total wall time used by the iterative GPR training and subsequent MCMC, using $6$ processors on a laptop, all the examples except $f_2$ completed in $\lesssim3$ minutes (in some cases, less than a minute), while $f_2$ required $\sim27$ minutes. For these simple examples, the ASA optimization step was dramatically faster in terms of wall time, simply because the cost functions we used could be calculated extremely fast. In fact, in almost all the above cases, the standard MCMC outperformed \textsc{picasa} in terms of wall time (especially for the non-linear $f_2$ example). The true utility of \textsc{picasa} is expected to emerge in problems where each cost function evaluation takes, say, tens of seconds or more on a single processor.\footnote{E.g., if the cost function of $f_2$ were to take 20 seconds for each evaluation and we used 16 processors, the standard MCMC shown in Fig.~\ref{fig:nonlin_result} would take $\sim75$ hours to complete $\sim217000$ evaluations. The ASA optimization in \textsc{picasa} with $\Nsamp=96$ (a multiple of 16) would use $N_{\rm eval}\sim5900$, taking $\sim2$ hours. Adding the $\sim0.5$ hours for GPR training, this still gives a factor $\sim30$ improvement in wall time. Upon increasing the computational expense of the cost function, eventually the GPR training ceases to be a significant fraction of the cost, and the relative wall time gain for \textsc{picasa} in this example would asymptote to the ratio of $N_{\rm eval}$ values, $\gtrsim35$.}

We chose to display 3$\sigma$ contours in the examples above, so as to assess the performance of \textsc{picasa} under stringent requirements of accurately reproducing the outer tails of the posterior distribution. If one is interested in at most 2$\sigma$ contours, \textsc{picasa} can be run with smaller values of $\delta_\sigma$ (data pack key \texttt{dsig}), leading to even fewer evaluations. Of course, in this case the standard MCMC would also require fewer evaluations, but we have checked that the relative improvements seen above do not change significantly. 
Finally, all our examples used diagonal covariance matrices for the data errors, so it is also interesting to ask whether correlated data errors have any impact on the performance of \textsc{picasa}. 
Using some simple forms for positive-definite non-diagonal covariance matrices $C_{\rm data}$ \citep[e.g.,][]{ms14-markov} to define data errors in the examples above, we have checked that the comparisons of $N_{\rm eval}$ between \textsc{picasa} and standard MCMC are essentially unchanged.

These examples demonstrate the basic behaviour of \textsc{picasa} in a variety of situations with different levels of non-linearity and dimensionality of parameter space. We discuss a few caveats and avenues for improvement in section~\ref{sec:conclude}. 

\section{Conclusion}
\label{sec:conclude}
We have presented a new technique for efficient parameter exploration, dubbed Anisotropic Simulated Annealing (ASA), particularly aimed at computationally expensive likelihoods. ASA builds on the ideas underlying simulated annealing and iteratively samples smaller and smaller regions of parameter space using Latin hypercubes, zooming in on the minimum of a pre-defined cost function. In addition to efficient recovery of the location of this minimum (i.e., the best-fit parameters), the resulting sample can be interpolated using Gaussian Process Regression to subsequently perform a standard MCMC algorithm with inexpensive sampling. We have combined these features into the Python code \textsc{picasa} ({\bf P}arameter {\bf I}nference using {\bf C}obaya with {\bf A}nisotropic {\bf S}imulated {\bf A}nnealing), with the integration with MCMC implemented using the \textsc{cobaya} framework \citep{tl19-cobaya,tl21-cobaya}. 

Through a number of linear and non-linear examples, we demonstrated that \textsc{picasa} can easily achieve effective gains in the number of likelihood evaluations of factors of $\gtrsim10$ to $\gtrsim100$, depending on the specific problem, in comparison to a standard MCMC approach. 
The code \textsc{picasa} is being actively developed and currently has several aspects which can be improved further. These include:
\begin{enumerate}
\item Handling situations where the best-fit lies close to a sharp prior (in other words, when the data provide a limit rather than a constraint on a parameter). Currently, the ASA optimization will find it troublesome to settle onto an acceptable fit in such cases, and the GPR training will also likely not converge well. It should be possible to address this issue by allowing for the likelihood to be evaluated (e.g., by analytic continuation) in undesirable regions of parameter space, and then separating the imposition of the prior from the likelihood evaluator, just like in standard MCMC implementations.
\item The efficiency and robustness of the GPR training can likely be enhanced by allowing for dynamic assessment of the best choice of kernel (akin to a low-level neural network), as well as automated and robust assessment of the need for additional $\chi^2$ samples. At present, the GPR training described in section~\ref{subsec:picasa:GPR} can require some level of user intervention, which is also not desirable.
\item Handling multiple (nearly) degenerate minima of the cost function. At present, \textsc{picasa} is guaranteed to remain stuck in one such minimum. It should be possible to allow for exploration of degenerate minima using, e.g., known or suspected symmetries of the cost function.
\item Handling non-Gaussian errors defining the log-likelihood. At present, \textsc{picasa} interprets the value of the cost function $\chi^2(\mathbf{a})$ assuming Gaussian errors on the observable data. This can be a problem when the likelihood is substantially non-Gaussian in the data (e.g., when analysing galaxy cluster counts for cosmological inference). Since the goodness-of-fit is only used to assess whether or not an acceptable minimum is being reached, at present we recommend shifting the definition of $\chi^2$ in such cases by an additive constant, to make its expected value near the best fit close to the number of degrees of freedom. It should, however, also be possible to modify this condition by allowing for alternate (more Bayesian, say) criteria to assess the quality of the fit.
\item Combining with approximate likelihood schemes. At present, \textsc{picasa} places the onus of accurate likelihood evaluation on the user. It will be interesting to explore whether approximation schemes such as DALI \citep{sqa14} can be integrated into \textsc{picasa} for increased robustness and speed, especially for estimating the best fit, GPR training and extensions to non-Gaussian data errors.
\end{enumerate}
These and other improvements will continue to be updated on the public repository for \textsc{picasa} listed under data availability below. We expect \textsc{picasa} to be useful in a number of situations requiring numerically expensive likelihood calls. A concrete example could be the reconstruction of the baryon acoustic oscillation (BAO) feature as discussed by \citet{nsz22}. More generally, since \textsc{picasa} is agnostic to the nature of the observable, the list of these potential applications can span numerous topics in cosmology, extra-Galactic astrophysics, gravitational waves science, exoplanetary science, and so on. We will explore some of these applications in future work.
%We conclude by listing a few of these here.
% \begin{itemize}
% \item \emph{Cosmology and extra-Galactic astrophysics}
% \begin{itemize}
% \item Voronoi reconstruction
% \item SCRIPT reionisation studies
% \item Lognormal Lyman alpha
% \item RC fitting
% \end{itemize}
% \item \emph{Gravitational wave physics}
% \begin{itemize}
% \item 
% \end{itemize}
% \item \emph{Exoplanet studies}
% \begin{itemize}
% \item 
% \end{itemize}
% \end{itemize}
% Although this list is restricted to problems in astrophysics and cosmology, the \textsc{picasa} framework is agnostic to the nature of the observable; this is evident from the examples in section~\ref{sec:examples}. As such, the utlity of \textsc{picasa} should extend to a wide variety of problems involving statistical inference.

\section*{Acknowledgments}
It is a pleasure to thank Tirthankar Roy Choudhury for encouragement and invaluable input in developing \textsc{picasa} functionality, and both him and Barun Maity for stress-testing \textsc{picasa}. I am also grateful to Ravi Sheth and Shadab Alam for collaboration on a project that used an early version of the ASA algorithm, as well as comments on an earlier draft. Finally, I thank Dipankar Bhattacharya for discussions regarding simulated annealing which originally sparked the ideas underlying this work.
This work made extensive use of the open source computing packages NumPy \citep{vanderwalt-numpy},\footnote{\url{http://www.numpy.org}} SciPy \citep{scipy},\footnote{\url{http://www.scipy.org}} Scikit-Learn \citep{scikit-learn},\footnote{\url{https://scikit-learn.org/}} Matplotlib \citep{hunter07_matplotlib}\footnote{\url{https://matplotlib.org/}} and Jupyter Notebook.\footnote{\url{https://jupyter.org}}

\section*{Data availability}
The code \textsc{picasa} is publicly available at \url{https://bitbucket.org/aparanjape/picasa/}. The \textsc{cobaya} framework for parameter inference is publicly available at \url{https://cobaya.readthedocs.io/}.

\bibliography{references}

\label{lastpage}
\end{document}